\documentclass[preprint]{ptephy_v1}

\usepackage{braket}
\usepackage{comment}

\preprintnumber{XXXX-XXXX} 

\def\Journal#1#2#3#4{{#1} {\bf #2}, #3 (#4)}
 
\def\ARNPS{Annu. Rev. Nucl. Part. Sci.} 
\def\AandA{Astron. Astrophys.} 

\def\APJ{Astrophys. J.}

\def\CMP{Commn. Math. Phys.}

\def\CPC{Chin. Phys. C}
\def\CQG{Class. Quant. Grav.}
\def\EPJC{Eur. Phys. J. C}
\def\EPJP{Eur. Phys. J. Plus}

\def\IJMPA{Int. J. Mod. Phys. A}
\def\IJMPD{Int. J. Mod. Phys. D}

\def\JCAP{J. Cosmol. Astropart. Phys.}
\def\JHEP{J. High Energy Phys.}

\def\MPLA{Mod. Phys. Lett. A}
\def\MNRAS{Mon. Not. R. Astron. Soc.}

\def\NATULONDON{Nature (London)}

\def\NPB{Nucl. Phys. B}

\def\PLB{{Phys. Lett.} B}

\def\PREP{Phys. Rep.}

\def\PRL{Phys. Rev. Lett.}
\def\PRD{Phys. Rev. D}

\def\PTP{Prog. Theor. Phys.}

\def\SCIENCE{Science}

\begin{document}

\title{Primordial black holes and lepton flavor violation with scotogenic dark matter}


\author{Teruyuki Kitabayashi}
\affil{Department of Physics, Tokai University,
4-1-1 Kitakaname, Hiratsuka, Kanagawa 259-1292, Japan\email{teruyuki@tokai-u.jp}}


\begin{abstract}
We show that if the lepton flavor violating $\mu \rightarrow e \gamma$ process is observed in the MEG II experiment, the initial density of primordial black holes (PBHs) can be constrained with the scotogenic dark matter. As a benchmark case, if the PBH evaporation occurs in the radiation dominated era, the initial density may be $2\times 10^{-17} \lesssim \beta \lesssim 3 \times 10^{-16}$ for $\mathcal{O}$ (TeV) scale dark sector in the scotogenic model where $\beta=\rho_{\rm PBH}/\rho_{\rm rad}$ is the ratio of the PBH density $\rho_{\rm PBH}$ to the radiation density $\rho_{\rm rad}$ at the time of PBH formation. As an other benchmark case, if PBHs evaporate in the PBH dominated era, the initial density may be $1 \times 10^{-8} \lesssim \beta \lesssim 3 \times 10^{-7}$ for $\mathcal{O}$ (GeV) scale dark matter with other $\mathcal{O}$ (TeV) scale particles in the scotogenic model.
\end{abstract}

\subjectindex{}

\maketitle

\section{Introduction\label{section:introduction}}
The primordial black holes (PBHs) are the one of the type of black hole that produced in the early Universe \cite{Carr1975APJ,Carr2010PRD,Carr2020ARNPS,Carr2020arXiv}. The PBHs are produced via a number of mechanisms, such as the collapse of large density perturbations generated from inflation \cite{Garcia-Bellido1996PRD,Kawasaki1998PRD,Yokoyama1998PRD,Kawasaki2006PRD,Kawaguchi2008MNRAS,Kohri2008JCAP,Drees2011JCAP,Lin2013PLB,Linde2013PRD}, a sudden reduction in the pressure \cite{Khlopov1980PLB,Jedamzik1997PRD}, bubble collisions \cite{Crawford1982NATULONDON,Hawking1982PRD,Kodama1982PTP,La1989PLB,Moss1994PRD}, a curvaton \cite{Yokoyama1997AandA,Kawasaki2013PRD,Kohri2013PRD,Bugaev2013IJMPD} and collapse of cosmic string \cite{Hogan1984PLB}. 

A PBH emits particles via the Hawking radiation \cite{Hawking1975CMP}. Since the Hawking radiation is induced by gravity, PBHs evaporate into all particle species. Thus, the study of PBH is important not only in the cosmology but also in the particle physics. For example, since the particle dark matter is also produced by Hawking radiation of the PBHs, the correlations between the initial density of the PBHs and dark matter mass have been extensively studied in the literature \cite{Bell1999PRD,Green1999PRD,Khlopov2006CQG,Baumann2007arXiv,Dai2009JCAP,Fujita2014PRD,Allahverdi2018PRD,Lennon2018JCAP,Morrison2019JCAP,Hooper2019JHEP,Masina2020EPJP,Baldes2020JCAP,Bernal2020arXiv1,Gondolo2020PRD,Bernal2020arXiv2,Auffinger2020arXiv,Datta2021arXiv,Chaudhuri202011arXiv,Kitabayashi2021IJMPA,Cheek2021arXiv1,Cheek2021arXiv2,Das2021arXiv}. For another example, the  influence of lepton flavor asymmetries on the mass spectrum of PBHs are recently studied \cite{Bodeker2021PRD,Vovchenko2021PRL}.

On the other hand, the lepton flavor violating phenomena, such as $\mu \rightarrow e \gamma$ process, are directly related to the new physics beyond the standard model of particle physics \cite{Lindner2018PhysRep}. The most of new physics models predict some lepton flavor violation effects. To test the availability of these new models, theoretical predictions of branching ratio of lepton flavor violating processes within these models are important. In addition, since the physics run of the MEG II experiment to search $\mu \rightarrow e \gamma$ process with 10 times better sensitivity than the MEG experiment will be started very near future \cite{Baldini2018EPJC,Ieki2020PoS}, a study which is related on the $\mu \rightarrow e \gamma$ process is interesting and timely. 

In this paper, we show the correlations between the initial density of PBHs and the branching ratio of $\mu \rightarrow e \gamma$ with scotogenic dark matter. The scotogenic dark matter \cite{Ma2006PRD} in the scotogenic model is one of the most successful and well studied dark matter candidate \cite{Suematsu2009PRD,Suematsu2010PRD,Kubo2006PLB,Hambye2007PRD,Farzan2009PRD,Farzan2010MPLA,Farzan2011IJMPA,Kanemura2011PRD,Schmidt2012PRD,Faezan2012PRD,Aoki2012PRD,Hehn2012PLB,Bhupal2012PRD,Bhupal2013PRD,Law2013JHEP,Kanemura2013PLB,Hirsch2013JHEP,Restrepo2013JHEP,Ho2013PRD,Lindner2014PRD,Okada2014PRD89,Okada2014PRD90,Brdar2014PLB,Toma2014JHEP,Ho2014PRD,Faisel2014PRD,Vicente2015JHEP,Borah2015PRD,Wang2015PRD,Fraser2016PRD,Adhikari2016PLB,Ma2016PLB,Arhrib2016JCAP,Okada2016PRD,Ahriche2016PLB,Lu2016JCAP,Cai2016JHEP,Ibarra2016PRD,Lindner2016PRD,Das2017PRD,Singirala2017CPC,Kitabayashi2017IJMPA,AbadaJHEP2018,Baumholzer2018JHEP,Ahriche2018PRD,Hugle2018PRD,Kitabayashi2018PRD,Reig2019PLB,Boer2020PRD,Ahriche2020PRD,Faisel2014PLB}. Since the scotogenic model can account for dark matter candidates and predict the lepton flavor violating processes simultaneously \cite{Suematsu2009PRD,Suematsu2010PRD}, and the scotogenic dark matter can be also produced by Hawking radiation of PBH \cite{Kitabayashi2021IJMPA}, the initial density of the PBHs and the branching ratio of $\mu \rightarrow e \gamma$ is related via scotogenic dark matter. 
 
The dark matter and PBH in the scotogenic model has already been discussed in Ref.\cite{Kitabayashi2021IJMPA} where the constraints from lepton flavor violating processes are also studied. We have to clear the difference between Ref.\cite{Kitabayashi2021IJMPA} and this paper. In this paper, we perform more advanced analysis by 
\begin{itemize}
\item takeing into account the expected sensitivity of the branching ratio of a lepton flavor violating $\mu \rightarrow e \gamma$ process from the future MEG II experiment (only current MEG constraint is taken into account in the Ref.\cite{Kitabayashi2021IJMPA}),
\item including the entropy production effect via the PBH evaporation into the numerical calculations in the PBH dominant case (there are only some comments about the effect of the entropy production in the Ref.\cite{Kitabayashi2021IJMPA}),
\item searching more wide parameter region of scotogenic mode (only TeV scale dark matter is considered as a typical value in the scotogenic model in the Ref.\cite{Kitabayashi2021IJMPA}).
\end{itemize}
Thanks to these new ingredients, especially including the expected results from the future MEG II experiment, the following new scientific findings are obtained in this paper:
\begin{itemize}
\item not only upper limit but also lower limit, allowed band, of the initial density of PBHs in the radiation dominant case (only upper limit of the initial density of PBHs are shown in the radiation dominant case in the Ref.\cite{Kitabayashi2021IJMPA}),
\item constraints of the initial density of PBHs in the PBH dominant case (there is no significant discussion for the constraint of the initial density of PBH in the PBH dominant case in the Ref.\cite{Kitabayashi2021IJMPA}).
\end{itemize}

This paper is organized as follows. In Sec. \ref{sec:scotogenic}, we present a review of the scotogenic model. In Sec. \ref{sec:LFVandPBH}, we show the correlations between the initial density of PBHs and the branching ratio of $\mu \rightarrow e \gamma$ with scotogenic dark matter. Section \ref{sec:summary} is devoted to a summary.

\section{Scotogenic model \label{sec:scotogenic}}
The scotogenic model \cite{Ma2006PRD} is an extension of the standard model in the particle physics. In this model,  three new Majorana $SU(2)_L$ singlets $N_k$ $(k=1,2,3)$ with mass $M_k$ and one new scalar $SU(2)_L$ doublet $(\eta^+,\eta^0)$ are introduced. These new particles are odd under exact $Z_2$ symmetry. The relevant Lagrangian and scalar potential for this paper are given by
\begin{eqnarray}
\mathcal{L} &=& Y_{\alpha k} (\bar{\nu}_{\alpha L} \eta^0 - \bar{\ell}_{\alpha L} \eta^+) N_k + \frac{1}{2}M_k \bar{N}_k N^C_k + h.c.,
\nonumber \\
V &=& \frac{1}{2}\lambda (\phi^\dagger \eta)^2 + h.c.,
\label{Eq:L_V}
\end{eqnarray}
where $L_\alpha=(\nu_\alpha, \ell_\alpha)$ $(\alpha=e,\mu,\tau)$ is the left-handed lepton doublet and $\phi=(\phi^+, \phi^0)$ is the standard Higgs doublet. 

Owing to the $Z_2$ symmetry, the tree level neutrino mass should vanish but they acquire masses via one-loop interactions. The flavor neutrino mass matrix $M$ is obtained as 
\begin{eqnarray}
M_{\alpha\beta} = \sum_{k=1}^3  \frac{\lambda v^2 Y_{\alpha k}Y_{\beta k}M_k}{16\pi^2(m^2_0-M^2_k)}\left(1-\frac{M^2_k}{m^2_0-M^2_k}\ln\frac{m_0^2}{M^2_k} \right),
\end{eqnarray}
where $m_0^2 = \frac{1}{2}(m_R^2+m_I^2)$ and $v$, $m_R$, $m_I$ denote vacuum expectation value of the Higgs field, the masses of $\sqrt{2} {\rm Re}[\eta^0]$ and $\sqrt{2} {\rm Im}[\eta^0]$, respectively. 

Since the lightest $Z_2$ odd particle is stable, it becomes a dark matter candidate. We assume that the lightest Majorana singlet fermion, $N_1$, is the dark matter particle. It is known that if the lightest singlet fermion is almost degenerated with the next to lightest singlet fermions, the observed relic abundance of dark matter $\Omega_{\rm DM} h^2 = 0.12 \pm 0.0009$ \cite{Planck2020AA} and the observed upper limit of the branching ratio of the $\mu \rightarrow e \gamma$ process ${\rm BR}(\mu\rightarrow e\gamma) \le 4.2\times 10^{-13}$ process \cite{MEG2016EPJC} can be simultaneously consistent with the prediction from the scotogenic model \cite{Suematsu2009PRD,Suematsu2010PRD}. Thus we set $M_1 \sim M_2 = 1.0001 M_1 < M_3<m_0$.

We would like to comment that if the DM and the new scalar particle are degenerate in mass, $M_1 \sim m_0$, their coannihilation processes become significant for the DM relic density. In this case, to compensate for DM relic density, the dark matter particle should be more heavier. The heavier dark matter yields more small ${\rm BR}(\mu\rightarrow e\gamma)$, for example, see Ref.\cite{Kitabayashi2021IJMPA}. Thus, if we include a parameter region with $M_1 \sim m_0$ in our numerical calculations, more stringent constraints of the parameters may be obtained. In this paper, we would like to keep the requirement of $M_1 < m_0$ and omit the possibility of the coannihilation of DM and new scalar particle. 

The relic abundance of cold dark matter which is produced by freeze-out mechanism is estimated to be \cite{Griest1991PRD}:
\begin{eqnarray}
\Omega_{\rm FO} h^2 = \frac{1.07\times 10^9 x_{\rm FO}}{g_\ast^{1/2} M_{\rm Pl} (a_{\rm eff}+3b_{\rm eff}/x_{\rm FO})},
\end{eqnarray}
where
\begin{eqnarray}
a_{\rm eff}= \frac{a_{11}}{4}+\frac{a_{12}}{2}+\frac{a_{22}}{4},  \quad
b_{\rm eff}= \frac{b_{11}}{4}+\frac{b_{12}}{2}+\frac{b_{22}}{4},
\label{Eq:aeff_beff}
\end{eqnarray}
with
\begin{eqnarray}
a_{ij}&=& \frac{1}{8\pi}\frac{M_1^2}{(M_1^2+m_0^2)^2} \sum_{\alpha\beta}(Y_{\alpha i} Y_{\beta j} - Y_{\alpha j} Y_{\beta i})^2, \nonumber \\
b_{ij}&=&\frac{m_0^4-3m_0^2M_1^2-M_1^4}{3(M_1^2+m_0^2)^2}a_{ij}  +  \frac{1}{12\pi}\frac{M_1^2(M_1^4+m_0^4)}{(M_1^2+m_0^2)^4}  \sum_{\alpha\beta}Y_{\alpha i} Y_{\alpha j} Y_{\beta i} Y_{\beta j},
\label{Eq:a_b}
\end{eqnarray}
and
\begin{eqnarray}
x_{\rm FO} = \frac{m_{\rm DM}}{T_{\rm FO}} \simeq 25,
\label{Eq:xf25}
\end{eqnarray}
is the freeze-out temperature \cite{KolbTurner1991}.

In this model, flavor violating processes such as $\mu \rightarrow e \gamma$ are induced at the one-loop level. The branching ratio of $\mu \rightarrow e \gamma$  is given by \cite{Kubo2006PLB}
\begin{eqnarray}
{\rm BR}(\mu \rightarrow e \gamma)=\frac{3\alpha_{\rm em}}{64\pi(G_{\rm F} m_0^2)^2}\left| \sum_{k=1}^3 Y_{\mu k}Y_{e k}^* F \left( \frac{M_k}{m_0}\right) \right|^2,
\end{eqnarray}
where $\alpha_{\rm em}$ denotes the fine-structure constant, $G_{\rm F}$ denotes the Fermi coupling constant and  $F(x)$ is defined by $F(x)=\frac{1-6x^2+3x^4+2x^6-6x^4 \ln x^2}{6(1-x^2)^4}$. 

The Yukawa coupling $Y$ can be express in terms of $\lambda, M_k, m_0$, neutrino masses $m_i$ ($i=1,2,3$), mixing angles $\theta_{ij}$ ($ij=12,23,13$), Dirac CP violating phase $\delta$ and Majorana CP phases $\alpha_i$ ($i=1,2$) by Casas-Ibarra parametrization \cite{Casas2001NPB}. Since the relic density of the scotogenic dark matter depends only weakly on CP-violating phases, the Majorana CP phases are neglected \cite{Boer2020PRD}.

We would like to comment that the contributions of Majorana phases would be small for the DM relic density but not for the neutrino phenomena. Since neutrino oscillation experiments are not sensitive to the Majorana phases, the neutrino phenomena with Majorana phases are often studied with cosmological discussion. For example, the so-called leptogenesis scenarios for the origin of the baryon asymmetry of the Universe \cite{Fukugita1986PLB} may be depend on the Majorana phases of neutrinos. In this paper, although we have attempted to obtain some connection between Majorana phases of neutrinos and PBH, since we just connect PBH and DM relic abundance, and DM relic abundance is not sensitive to the Majorana phases, we can not any prediction for Majorana phases with PBH at present. Up to now, the leptogenesis with scotogenic dark matter with PBH \cite{Das2021arXiv} and the interplay between thermal and PBH induced leptogenesis \cite{Perez-Gonzalez2021PRD} have been studied; however there is no significant prediction for Majorana phases with PBH. Some studies about the relation between Majorana phases of neutrinos and PBH are required in the future. 

We use the best-fit values of neutrino parameters in Ref. \cite{Esteban2020JHEP}. For simplicity, we assume the normal mass ordering for the neutrinos $m_1 < m_2 < m_3$. According to constraint $\sum m_i < 0.12 - 0.69$ eV from observation of CMB radiation \cite{Planck2020AA,Capozzi2020PRD}, we require 
\begin{eqnarray}
\sum m_i < 0.12 \ {\rm eV},
\end{eqnarray}
and 
\begin{eqnarray}
m_1= 0.001 - 0.1 \ {\rm eV}.
\end{eqnarray}
In addition, we have an observed upper limit $|M_{ee}| < 0.066 - 0.155$ eV from neutrinoless double beta decay experiment \cite{Capozzi2020PRD,GERDA2019Science}. We require a condition 
\begin{eqnarray}
|M_{ee}| < 0.066 \ {\rm eV},
\end{eqnarray}
in our numerical calculations.

For remaining 4 model parameters, $\{M_1, M_3, m_0, \lambda\}$, we set the commonly used following range \cite{Kubo2006PLB,Ibarra2016PRD,Lindner2016PRD}:  
\begin{eqnarray}
&& 1.0  \ {\rm GeV} \  \le M_1, M_3, m_0 \le 1.0 \times10^6 \ {\rm GeV}, \nonumber \\
&& 1.0 \times 10^{-11} \le \lambda \le 1.0 \times 10^{-6},
\end{eqnarray}
in our numerical calculations. Since we assume that the lightest Majorana fermion $N_1$ is the dark matter particle, hereafter we use a notation of $m_{\rm DM} = M_1$.

\section {Initial density of primordial black holes \label{sec:LFVandPBH}}

\subsection{Primordial black holes \label{sec:PBH}}

We assume that PBHs are produced in the early Universe by large density perturbations generated from an inflation \cite{Garcia-Bellido1996PRD,Kawasaki1998PRD,Yokoyama1998PRD,Kawasaki2006PRD,Kawaguchi2008MNRAS,Kohri2008JCAP,Drees2011JCAP,Lin2013PLB,Linde2013PRD} and a PBH's mass is proportional to a horizon mass and that PBHs have the same masses at their formation time. In addition, we assume that the PBHs form during the radiation dominated era, with a monochromatic mass function.

The temperature of the Universe at PBH formation time is obtained as 
\begin{eqnarray}
T_{\rm in} = \frac{\sqrt{3}5^{1/4}}{2\pi^{3/4}}\frac{\gamma^{1/2}}{g_*(T_{\rm in})^{1/4}}\left( \frac{M_{\rm Pl}^3}{M_{\rm in}} \right)^{1/2},
\label{Eq:Tin}
\end{eqnarray}
where $\gamma \sim 0.2$ \cite{Carr1975APJ}, $g_\ast$ is the relativistic effective degrees of freedom for the radiation energy density, $M_{\rm Pl} \simeq 1.221 \times 10^{19}$ GeV is the Planck mass and $M_{\rm in}$ is the initial mass of PBH.

We introduce the dimensionless parameter
\begin{eqnarray}
\beta = \frac{\rho_{\rm PBH}(T_{\rm in})}{\rho_{\rm rad}(T_{\rm in})},
\label{Eq:beta}
\end{eqnarray}
to represent the initial energy density of PBHs at the time of its formation, $\rho_{\rm PBH}(T_{\rm in})$, where $\rho_{\rm rad}(T_{\rm in})$ is the energy density of radiation.

A black hole loses its mass by producing particles with masses below the Hawking temperature
\begin{eqnarray}
T_{\rm BH} = \frac{M_{\rm Pl}^2}{8\pi M_{\rm BH}},
\end{eqnarray}
via Hawking radiation \cite{Hawking1975CMP}. Ignoring gray body factors, the energy spectrum of the Hawking radiation is similar to the Planck distribution (The effect of the gray body factor in the high energy geometrical optics limit are shown in Refs. \cite{Carr2010PRD,Lennon2018JCAP,Baldes2020JCAP,Auffinger2020arXiv}). The temperature of the Universe right after PBH evaporation is
\begin{eqnarray}
T_{\rm evap} = \frac{\sqrt{3} g_* (T_{\rm BH})^{1/4}}{64 \sqrt{2}5^{1/4}\pi^{5/4}} \left( \frac{M_{\rm Pl}^5}{M_{\rm in}^3}\right)^{1/2}.
\label{Eq:Tevap}
\end{eqnarray}

The PBHs emit scotogenic dark matter via the Hawking radiation \cite{Kitabayashi2021IJMPA}. Gondolo et al. \cite{Gondolo2020PRD} show that if PBH evaporate after the freeze-out of the dark matter, $T_{\rm FO} > T_{\rm evap}$, then the dark matter particles produced from PBHs may contribute to the final relic abundance of the dark matter. The criteria $T_{\rm FO} > T_{\rm evap}$ is translated into 
\begin{eqnarray}
\frac{M_{\rm in}}{M_{\rm Pl}} \gtrsim 2 \times 10^{12} \left( \frac{{\rm GeV}}{m_{\rm DM}} \right)^{2/3},
\end{eqnarray}
by Eqs. (\ref{Eq:xf25}) and (\ref{Eq:Tevap}). Since PBHs should be evaporated before big bang nucleosynthesis \cite{Fujita2014PRD,Kohri1999PRD}, the upper limit $M_{\rm in} \lesssim 1 \times 10^9$ g ($ M_{\rm in}/M_{\rm Pl} \lesssim 4.6 \times 10^{13}$) is obtained \cite{Carr2020arXiv}. We set conservatively the upper and lower bound of the initial PBH mass as follows:
\begin{eqnarray}
2 \times 10^{12} \left( \frac{{\rm GeV}}{m_{\rm DM}} \right)^{2/3} \le \frac{M_{\rm in}}{M_{\rm Pl}} \le 2 \times 10^{13}.
\end{eqnarray}
%

\begin{figure}[t]
\begin{center}
\includegraphics{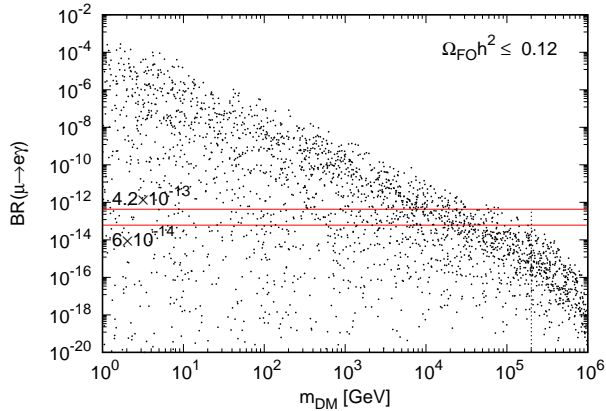}
\caption{Prediction of ${\rm BR}(\mu\rightarrow e\gamma)$ for $\Omega_{\rm FO} h^2 \le 0.12$ in the scotogenic model. The upper horizontal line shows the current observed upper limit of ${\rm BR}(\mu\rightarrow e\gamma) \le 4.2\times 10^{-13}$ from MEG experiment. The lower horizontal line shows the expected sensitivity of the future MEG II experiment: ${\rm BR}(\mu\rightarrow e\gamma) \simeq 6 \times 10^{-14}$. }
\label{fig:scotogenic}
\end{center}
\end{figure}

\begin{figure}[t]
\begin{center}
\includegraphics{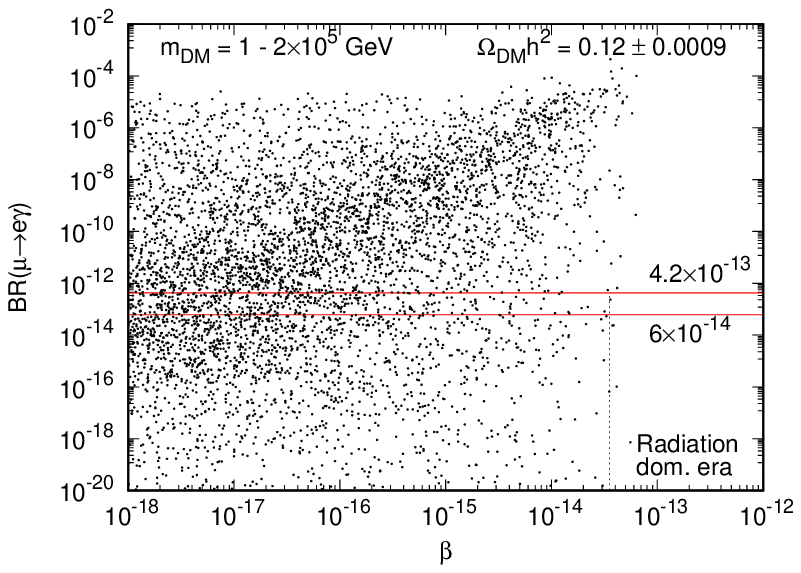}
\caption{Correlations between the branching ratio ${\rm BR} (\mu \rightarrow e \gamma)$ and initial density of PBHs $\beta$ for $\Omega_{\rm DM} h^2 = 0.12 \pm 0.0009$ in the case of $\beta < \beta_{\rm c}$ (radiation dominated era).  The upper horizontal line shows the current observed upper limit of ${\rm BR}(\mu\rightarrow e\gamma)$. The lower horizontal line shows the expected sensitivity of the future MEG II experiment.}
\label{fig:RD}
\end{center}
\end{figure}
\begin{figure}[t]
\begin{center}
\includegraphics[scale=0.8]{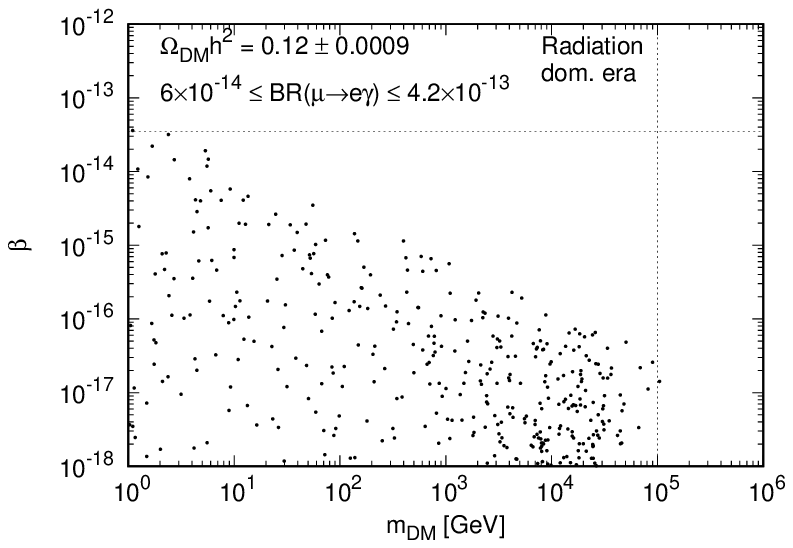}
\includegraphics[scale=0.8]{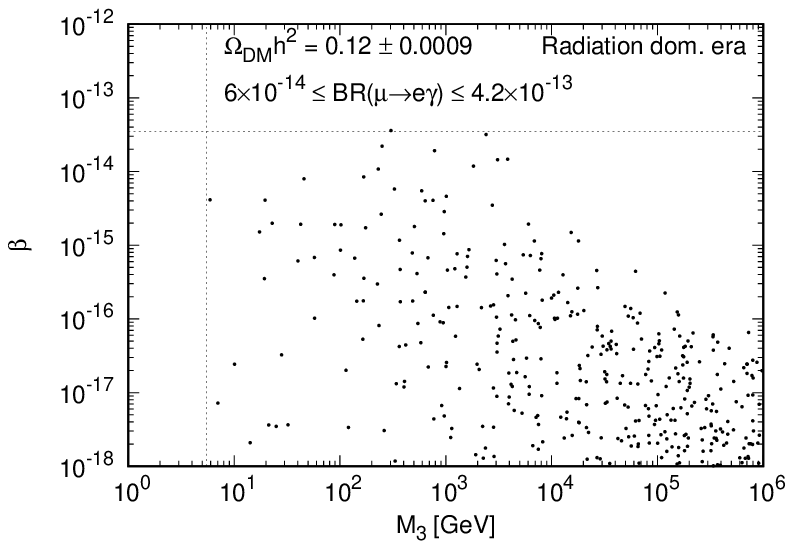}\\
\includegraphics[scale=0.8]{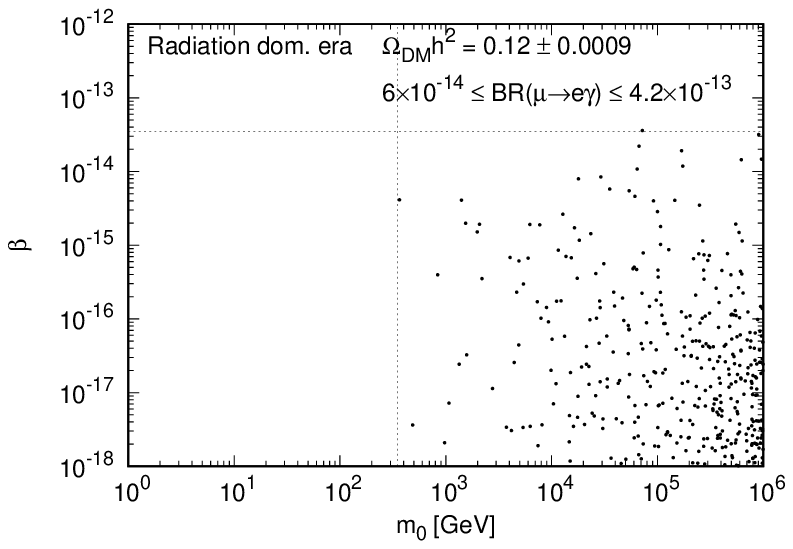}
\includegraphics[scale=0.8]{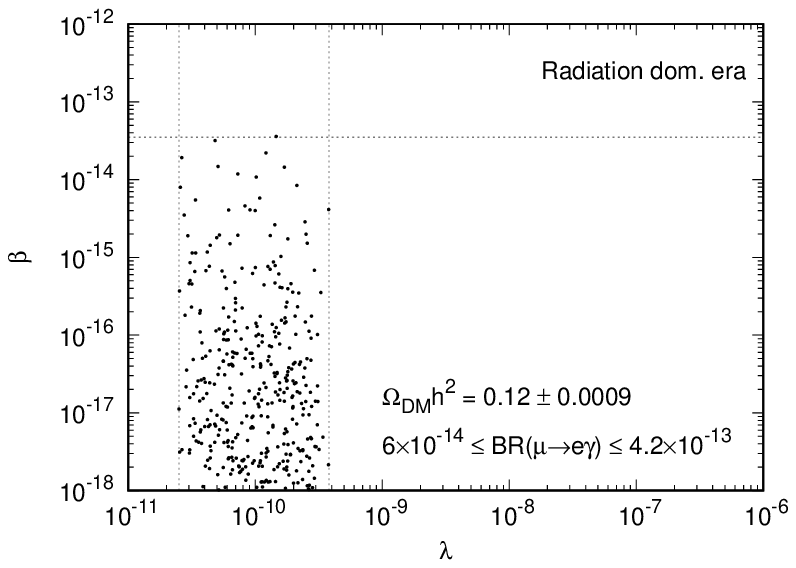}
\caption{Correlations between the initial density of PBHs $\beta$ and 4 free parameters $\{m_{\rm DM}(=M_1), M_3, m_0, \lambda\}$ in the scotogenic model for $\Omega_{\rm DM}h^2 = 0.12 \pm 0.0009$ and $6 \times 10^{-14} \le {\rm BR}(\mu \rightarrow e \gamma) \le 4.2\times 10^{-13}$ in the case of $\beta < \beta_{\rm c}$ (radiation dominated era).}
\label{fig:RD_beta_param}
\end{center}
\end{figure}

\begin{figure}[t]
\begin{center}
\includegraphics[scale=0.8]{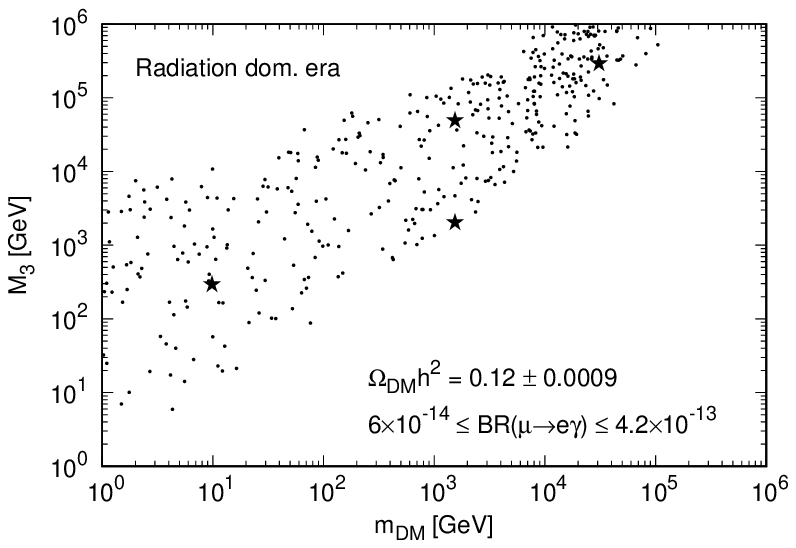}
\includegraphics[scale=0.8]{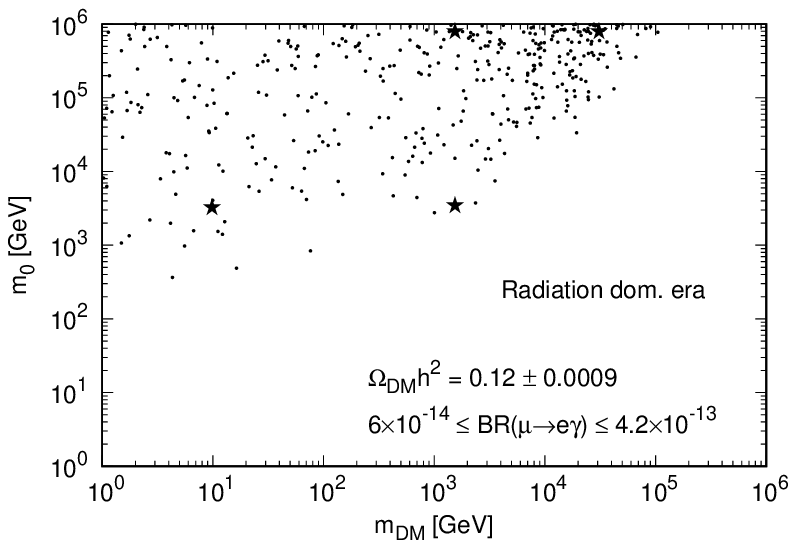}\\
\includegraphics[scale=0.8]{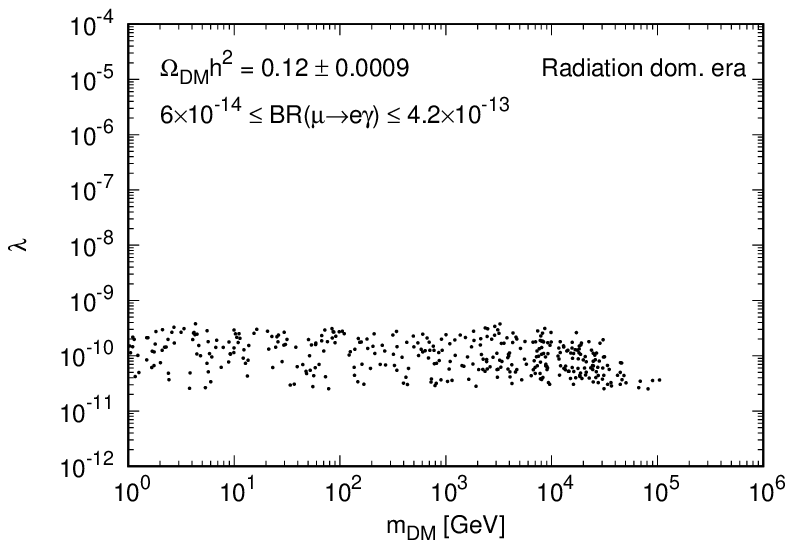}
\includegraphics[scale=0.8]{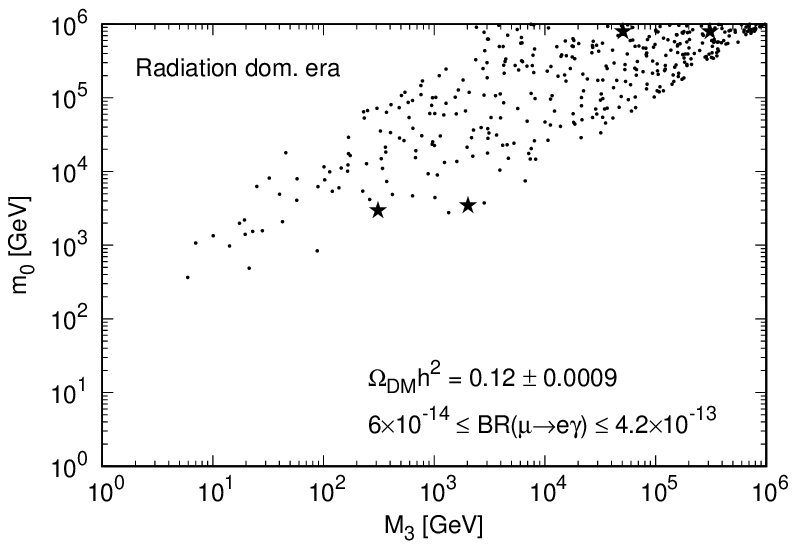}\\
\includegraphics[scale=0.8]{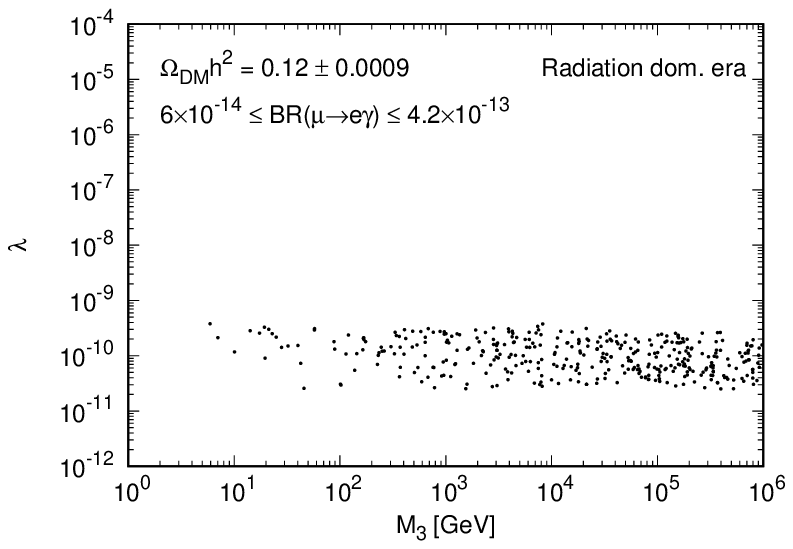}
\includegraphics[scale=0.8]{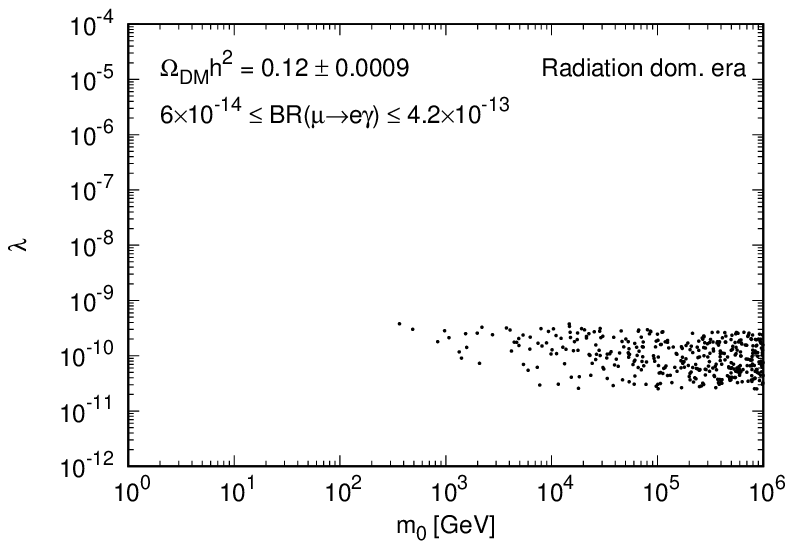}
\caption{Correlations between the allowed regions of 4 free parameters $\{m_{\rm DM}(=M_1), M_3, m_0, \lambda\}$ in the scotogenic model for $\Omega_{\rm DM}h^2 = 0.12 \pm 0.0009$ and $6 \times 10^{-14} \le {\rm BR}(\mu \rightarrow e \gamma) \le 4.2\times 10^{-13}$ in the case of $\beta < \beta_{\rm c}$ (radiation dominated era). The mark $\star$ denotes a benchmark case.}
\label{fig:RD_socoto_params}
\end{center}
\end{figure}
\begin{figure}[t]
\begin{center}
\includegraphics[scale=0.8]{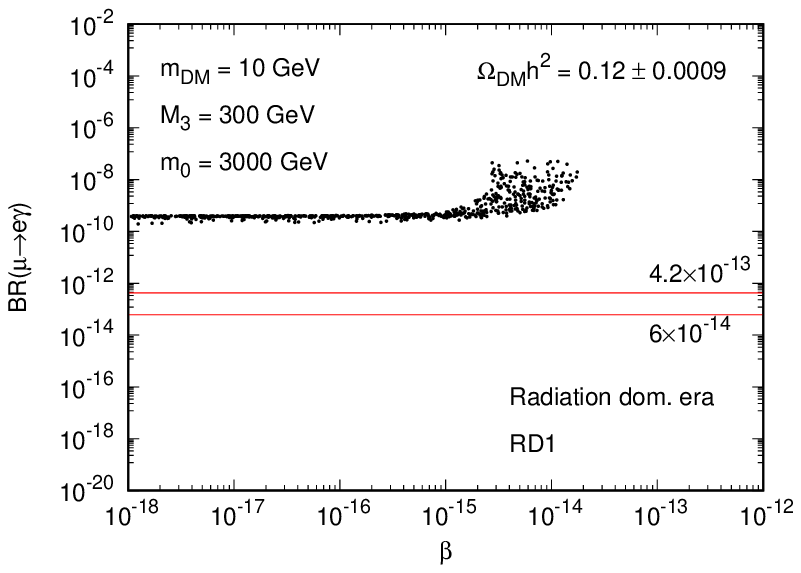}
\includegraphics[scale=0.8]{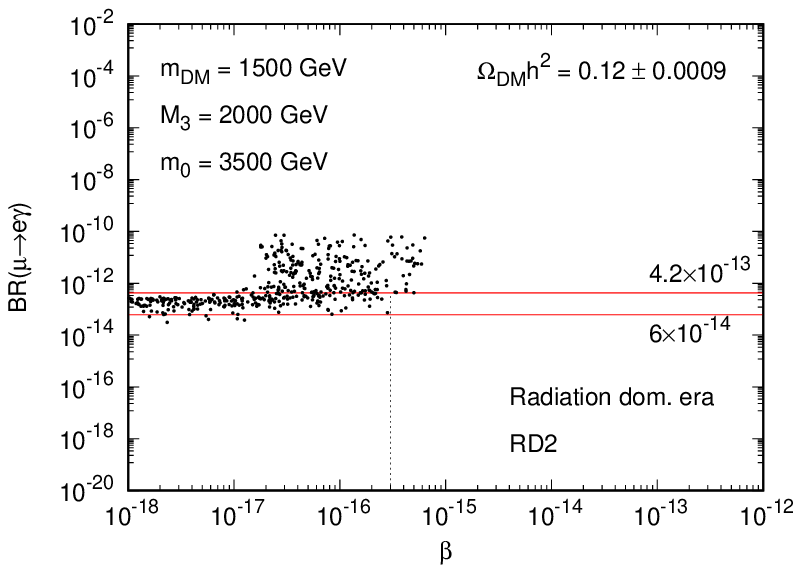}\\
\includegraphics[scale=0.8]{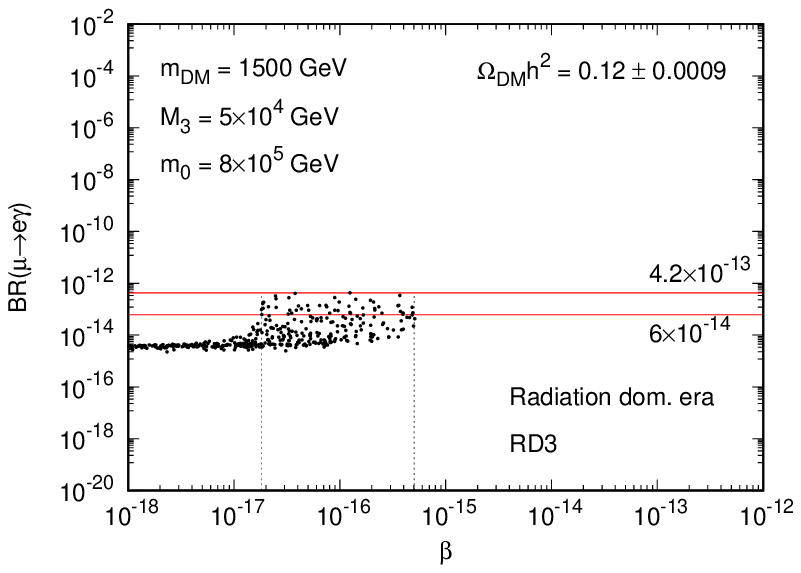}
\includegraphics[scale=0.8]{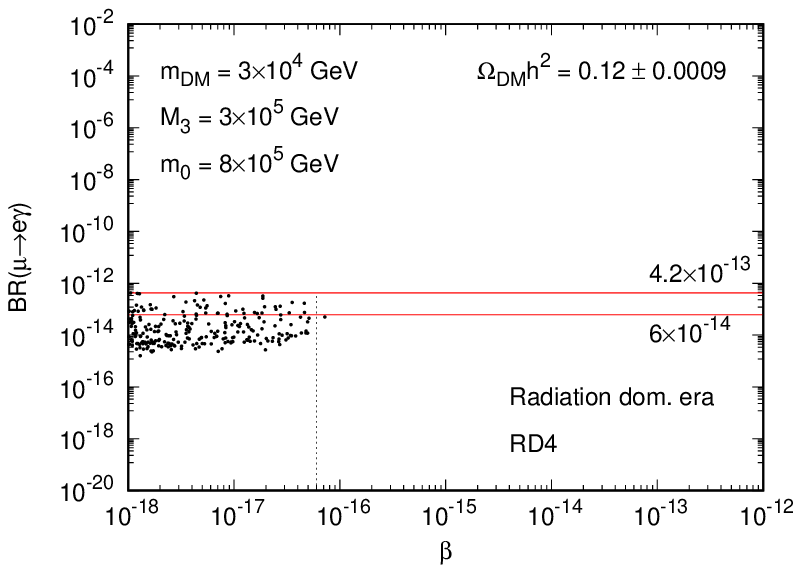}
\caption{Same of Fig. \ref{fig:RD} but for the benchmark cases.}
\label{fig:RD_BP}
\end{center}
\end{figure}

\subsection{PBH evaporation in radiation dominated era \label{sec:RD}}
Since $\rho_{\rm PBH} \propto a^{-3}$ and $\rho_{\rm rad}\propto a^{-4}$, where $a$ denotes the scale factor, $\rho_{\rm PBH}(t_{\rm early-eq}) \simeq \rho_{\rm rad}(t_{\rm early-eq})$ may be happen at the early equality time $t_{\rm early-eq}$. In order for PBH evaporation to occur before the early equality time  (radiation dominated era), $t_{\rm evap} < t_{\rm early-eq}$, the initial density of PBHs should be less than the following critical density ($\beta < \beta_{\rm c}$) \cite{Fujita2014PRD,Hamdan2018MPLA,Masina2020EPJP,Baldes2020JCAP}:
\begin{eqnarray}
\beta_{\rm c} &=& \frac{T_{\rm evap}}{T_{\rm in}} \nonumber \\
 &=& \sqrt{\frac{g_*(T_{\rm BH})}{10240\pi\gamma}}\left(\frac{M_{\rm in}} {M_{\rm Pl}}\right)^{-1} \label{Eq:RDconstraint} \\
&=& 0.129 \left(\frac{g_*(T_{\rm BH})}{106.75}\right)^{1/2} \left(\frac{0.2}{\gamma}\right)^{1/2}\left(\frac{M_{\rm in}} {M_{\rm Pl}}\right)^{-1}. \nonumber
\label{Eq:beta_c}
\end{eqnarray}

In this case, the final relic abundance of the scotogenic dark matter is to be
\begin{eqnarray}
\Omega_{\rm DM}h^2 = \Omega_{\rm FO}h^2 + \Omega_{\rm PBH}h^2,
\end{eqnarray}
where $\Omega_{\rm PBH}h^2$ denotes the relic abundance of scotogenic dark matter which is generated by PBH evaporation. 

At least, the relic abundance via freeze-out mechanism, $\Omega_{\rm FO}h^2$, should be less than the observed relic abundance of dark matter: $\Omega_{\rm DM} h^2 = 0.12 \pm 0.0009$. 

First, we estimate the allowed region of the dark matter mass $m_{\rm DM}$ without effect of the PBH evaporation. Figure \ref{fig:scotogenic} shows the prediction of ${\rm BR}(\mu\rightarrow e\gamma)$ for $\Omega_{\rm FO} h^2 \le 0.12 \pm 0.0009$ in the scotogenic model. The upper horizontal line shows the current observed upper limit of ${\rm BR}(\mu\rightarrow e\gamma) \le 4.2\times 10^{-13}$ from MEG experiment \cite{MEG2016EPJC}. The lower horizontal line shows the expected sensitivity of the future MEG II experiment: ${\rm BR}(\mu\rightarrow e\gamma) \simeq 6 \times 10^{-14}$ \cite{Baldini2018EPJC,Ieki2020PoS}. We observe that the allowed mass of scotogenic dark matter is  constrained with the observed upper limits of ${\rm BR}(\mu \rightarrow e \gamma)$ from MEG and MEG II experiments in the case of $\beta < \beta_{\rm c}$. From Fig. \ref{fig:scotogenic}, we perform our numerical studies in the mass region of 
\begin{eqnarray}
1 \ {\rm GeV} \le m_{\rm DM} \le 2 \times 10^{5} \ {\rm GeV},
\end{eqnarray}
in the case of $\beta < \beta_{\rm c}$.

We note that the upper limit of lepton flavor violating $\tau\rightarrow \mu\gamma$ and $\tau\rightarrow e\gamma$ processes are also measured as ${\rm BR}(\tau\rightarrow \mu\gamma) \le 4.4\times 10^{-8}$ and ${\rm BR}(\tau\rightarrow e\gamma) \le 3.3\times 10^{-8}$ \cite{BABAR2010PRL}; however, we only account for ${\rm Br}(\mu\rightarrow e\gamma)$ since it is the most stringent constraint. 

Now, we include the effect of the PBH evaporation in our analysis. For $1 \ {\rm GeV} \le m_{\rm DM} \le 2 \times 10^{5} \ {\rm GeV}$, the initial PBH mass should be
\begin{eqnarray}
\frac{M_{\rm in}}{M_{\rm Pl}} \gtrsim 
\begin{cases}
5.84 \times 10^{8} & (m_{\rm DM} = 2 \times 10^{5} {\rm GeV})\\ 
2 \times 10^{12} & (m_{\rm DM} = 1 \ {\rm GeV})
\end{cases},
\end{eqnarray}
for $T_{\rm FO} > T_{\rm evap}$. The corresponding Hawking temperature at the PBH formation time to be
\begin{eqnarray}
T_{\rm BH}^{\rm in} = 
\begin{cases}
8.3 \times 10^{8} {\rm GeV} & (\frac{M_{\rm in}}{M_{\rm Pl}} = 5.84 \times 10^{8}) \\ 
2.4 \times 10^{5} {\rm GeV} & (\frac{M_{\rm in}}{M_{\rm Pl}} = 2 \times 10^{12}) 
\end{cases}.
\end{eqnarray}
Thus, the relation $T_{\rm BH}^{\rm in} > m_{\rm DM}$ is satisfied in our setup. The relic abundance of scotogenic dark matter which is generated by PBH evaporation is obtained as
\begin{eqnarray}
\Omega_{\rm PBH}  h^2 &\simeq&  7.31 \times 10^7  \left(\frac{g_*(T_{\rm in})}{106.75} \right)^{-1/4}  
 \beta  \frac{3}{4} \frac{g_{\rm DM}  }{g_*(T_{\rm BH})}   \left( \frac{m_{\rm DM}}{{\rm GeV}} \right) \left( \frac{M_{\rm in}}{M_{\rm Pl}} \right)^{1/2},\label{Eq:OmegaPBHh2_RD_T>m} 
\end{eqnarray}
for $T_{\rm BH}^{\rm in} > m_{\rm DM}$ \cite{Fujita2014PRD,Hamdan2018MPLA,Masina2020EPJP,Baldes2020JCAP}. 

Figure \ref{fig:RD} shows the correlations between the branching ratio ${\rm BR} (\mu \rightarrow e \gamma)$ and initial density of PBHs $\beta$ for observed relic abundance of dark matter $\Omega_{\rm DM} h^2 = 0.12 \pm 0.0009$ \cite{Planck2020AA} in the case of $\beta < \beta_{\rm c}$. The upper horizontal line shows the current observed upper limit of ${\rm BR}(\mu\rightarrow e\gamma)$. The lower horizontal line shows the expected sensitivity of the future MEG II experiment. From Fig. \ref{fig:RD}, the initial density of PBHs should be 
\begin{eqnarray}
\beta \lesssim 3.5 \times 10^{-14},
\end{eqnarray}
for $\Omega_{\rm DM} h^2 = 0.12 \pm 0.0009$ in the case of  $\beta < \beta_{\rm c}$ with scotogenic dark matter. 

Since we have 4 free parameters $\{m_{\rm DM}(=M_1), M_3, m_0, \lambda\}$ within wide range, a deeper numerical study around these 4 free parameters is necessary. 

Figure \ref{fig:RD_beta_param} shows the correlations between the initial density of PBHs $\beta$ and 4 free parameters $\{m_{\rm DM}(=M_1), M_3, m_0, \lambda\}$ in the scotogenic model for  $\Omega_{\rm DM}h^2 = 0.12 \pm 0.0009$ and $6 \times 10^{-14} \le {\rm BR}(\mu\rightarrow e\gamma) \le 4.2\times 10^{-13}$ in the case of $\beta < \beta_{\rm c}$. From Fig. \ref{fig:RD_beta_param}, if $\mu \rightarrow e \gamma$ process is observed in the MEG II experiment, the allowed regions of the 4 free parameters to be:
\begin{eqnarray}
1.0 \lesssim &m_{\rm DM} \ [{\rm GeV}]&  \lesssim 1 \times 10^{5}, \nonumber \\
5.5  \lesssim &M_3 \ [{\rm GeV}]& \lesssim 1 \times 10^{6}, \nonumber \\
3.5 \times 10^{2} \lesssim &m_0 \ [{\rm GeV}]&  \lesssim  1 \times 10^{6}, \nonumber \\
2.5 \times 10^{-11} \lesssim &\lambda&  \lesssim 3.8 \times 10^{-10}, 
\end{eqnarray}
in the case of $\beta < \beta_{\rm c}$. The magnitude of $\lambda$ is constrained around $10^{-10}$ for $\beta < \beta_{\rm c}$. In the case of $\beta < \beta_{\rm c}$ (radiation dominated era), the relic abundance via freeze-out mechanism within the scotogenic model may be more dominant than the relic abundance via PBH evaporation in the observed relic abundance \cite{Gondolo2020PRD}. Thus, this characteristic constraint of $\lambda$ is needed to satisfy the experimental constraint, especially ${\rm BR}(\mu\rightarrow e\gamma)$, in the scotogenic model. On the other hand, as we show later, the relic abundance via PBH evaporation becomes more dominant in the observed relic abundance in the case of $\beta > \beta_{\rm c}$ (PBH dominated era) and more wide parameter region of $\lambda$ becomes to be allowed. 

Here we would like to show the additional discussions on the structure of Yukawa couplings to understand the phenomenology in the model. As shown in Refs, \cite{Vicente2015JHEP,Boer2020PRD}, the small $\lambda_5$, $\lambda_5 \sim 10^{-10}$, yields relatively large Yukawa couplings, e.g., $|y_1| = (0.078)\pm 0.021 \sqrt{m_{\rm DM}/{\rm GeV}}$ with $\lambda_5 \sim 10^{-10}$ for normal ordering neutrino mass may be expected where $y_1$ denotes an eigenvalue of the Yukawa matrix \cite{Boer2020PRD} and the typical magnitude of the Yukawa couplings $|Y_{\alpha k}|$ may be $\mathcal{O}$(0.1-1) \cite{Vicente2015JHEP}. Since the Yukawa couplings become relatively large, the cancelation among the Yukawa couplings needed to satisfy the experimental constraints. For example, the relatively large Yukawa couplings $Y_{e1}=0.77$, $Y_{e2}=-0.25-0.026i$, $Y_{e3}=0.46-0.022i$, $Y_{\mu 1}=0.52$, $Y_{\mu 2}=0.58-0.017i$, $Y_{\mu 3}=-0.55-0.015i$, $Y_{\tau 1}=-0.18-0.056i$, $Y_{\tau 2}=0.97$, $Y_{\tau 3}=0.84$ are obtained for $\lambda = 10^{-10}$ in our numerical calculations; however, the cancelation among these Yukawa couplings yields $|\sum_{k=1}^3 Y_{\mu k} Y^\ast_{e k} F(M_k/m_0)|^2 = 4.4\times 10^{-7}$ and we can obtain an acceptable small magnitude of BR($\mu \rightarrow e\gamma$) = $1.2 \times 10^{-13}$.

Figure \ref{fig:RD_socoto_params} shows the correlations between the allowed regions of 4 free parameters $\{m_{\rm DM}(=M_1), M_3, m_0, \lambda\}$ in the scotogenic model in the case of $\beta < \beta_{\rm c}$. We pick up the following 4 benchmark cases in the case of $\beta < \beta_{\rm c}$ (radiation dominated era (RD)):
\begin{description}
\item[RD1:] $\{m_{\rm DM}, M_3, m_0\} =\{10, 300, 3000 \}$ GeV as a light masses set.
\item[RD2:] $\{m_{\rm DM}, M_3, m_0\} =\{1500, 2000, 3500\}$ GeV as a middle masses set. The energy scale in this benchmark case could be target of research in the next generation experiments \cite{Kubo2006PLB,Ibarra2016PRD,Lindner2016PRD}. 
\item[RD3:] $\{m_{\rm DM}, M_3, m_0\} =\{1500, 5\times 10^4, 8\times 10^5\}$ GeV  as an other middle masses set.
\item[RD4:] $\{m_{\rm DM}, M_3, m_0\} =\{3\times 10^4, 3\times 10^5, 8\times 10^5 \}$ GeV  as a heavy masses set.
\end{description}
The mark $\star$ denotes a benchmark case in Fig. \ref{fig:RD_socoto_params}. To avoid the large number of benchmark case, we have distinguished the benchmark cases by 3 parameter $\{m_{\rm DM}, M_3, m_0\}$ without $\lambda$. We take any value of $\lambda$ in it's allowed region.

Figure \ref{fig:RD_BP} shows the same of Fig. \ref{fig:RD} but for the benchmark cases. We observe that if the lepton flavor violating $\mu \rightarrow e \gamma$ process is observed in the MEG II experiment, $4.2 \times 10^{-13} \lesssim {\rm BR}(\mu\rightarrow e\gamma) \lesssim 6 \times 10^{-14}$, the initial density of PBHs should be constrained for each benchmark case as follows:
\begin{description}
\item[RD1:] No constraint.
\item[RD2:] $\beta \lesssim 3 \times 10^{-16}$.
\item[RD3:] $2 \times 10^{-17}  \lesssim \beta \lesssim 5 \times 10^{-16}$.
\item[RD4:] $\beta \lesssim 6 \times 10^{-17}$.
\end{description}
%

\subsection{PBH evaporation in PBH dominated era \label{sec:PBH-dom}}

\begin{figure}[t]
\begin{center}
\includegraphics{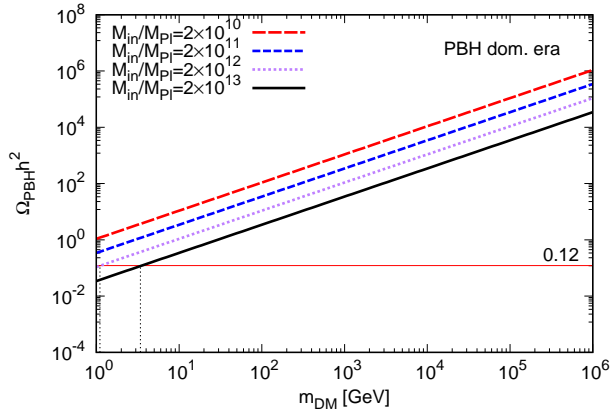}
\caption{Dependence of relic abundance of PBH-origin scotogenic dark matter  $\Omega_{\rm PBH}h^2$ on dark matter mass $m_{\rm DM}$ in the case of $\beta > \beta_{\rm c}$ (PBH dominated era). The horizontal line shows the observed relic abundance of dark matter.}
\label{fig:PBHdom_omegaPBH_mDM}
\end{center}
\end{figure}

\begin{figure}[t]
\begin{center}
\includegraphics{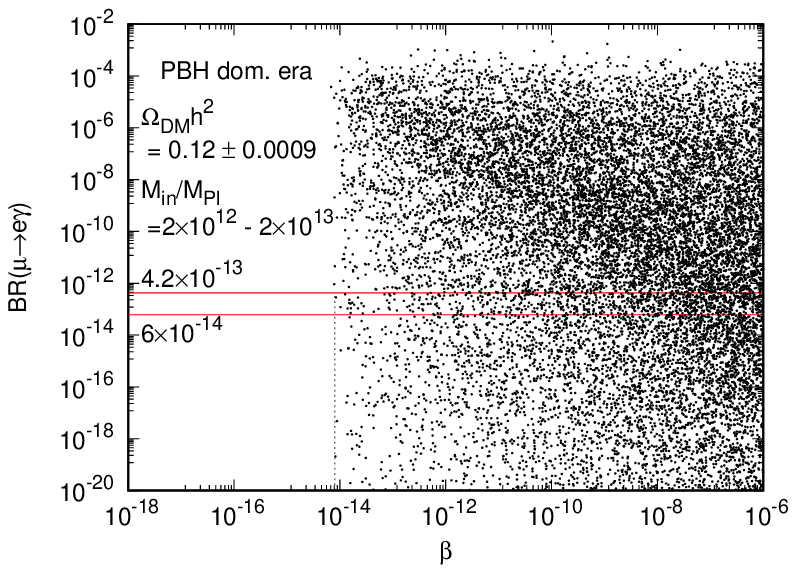}
\caption{Correlations between the branching ratio ${\rm BR} (\mu \rightarrow e \gamma)$ and initial density of PBHs $\beta$ for $\Omega_{\rm DM} h^2= 0.12 \pm 0.0009$ in the case of $\beta > \beta_{\rm c}$ (PBH dominated era). The upper horizontal line shows the current observed upper limit of ${\rm BR}(\mu\rightarrow e\gamma)$. The lower horizontal line shows the expected sensitivity in the future MEG II experiment.}
\label{fig:PBHdom_Br_beta}
\end{center}
\end{figure}

\begin{figure}[t]
\begin{center}
\includegraphics[scale=0.8]{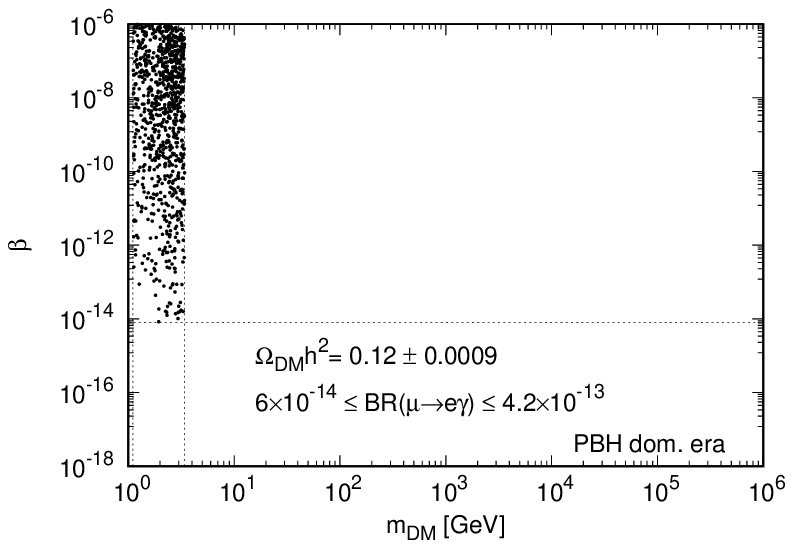}
\includegraphics[scale=0.8]{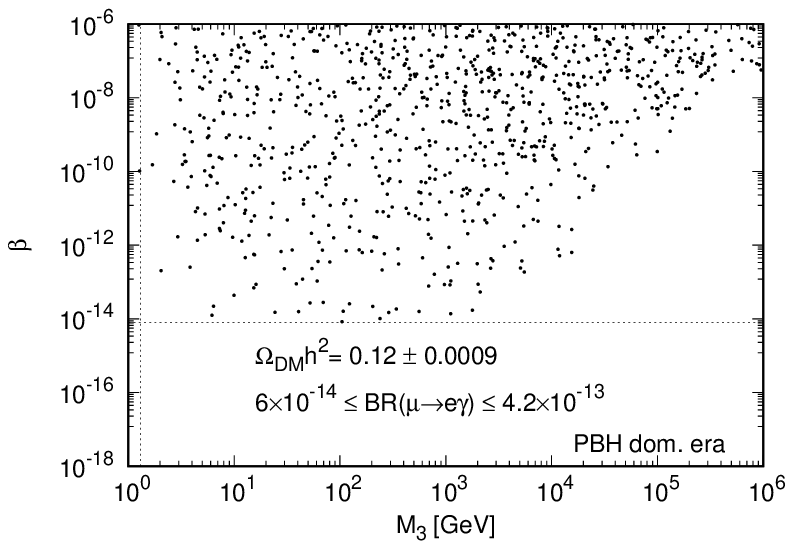}\\
\includegraphics[scale=0.8]{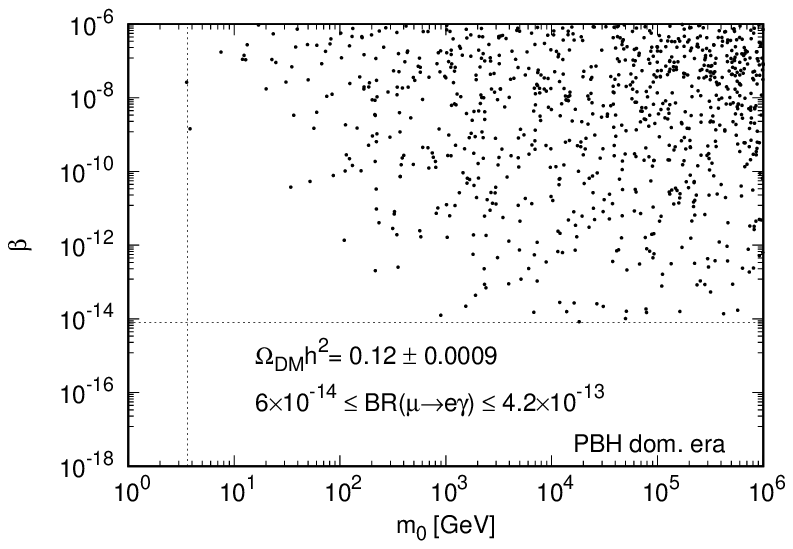}
\includegraphics[scale=0.8]{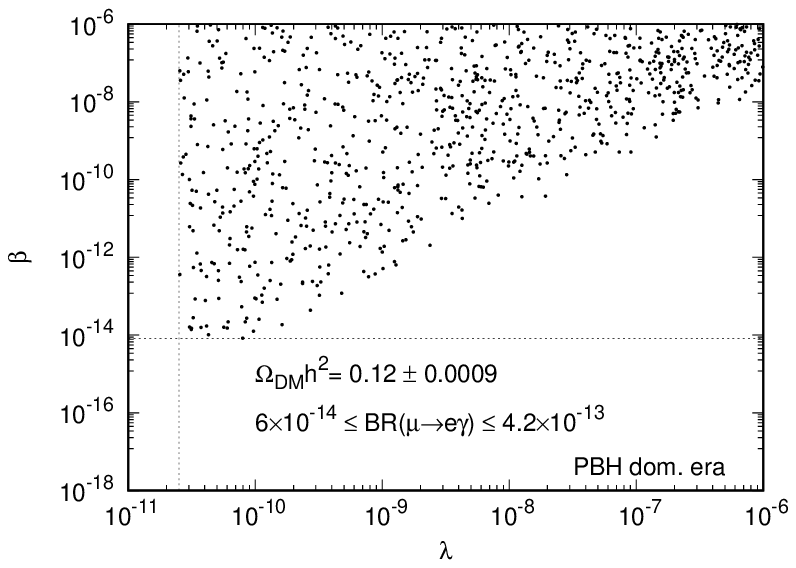}
\caption{Correlations between the initial density of PBHs $\beta$ and 4 free parameters $\{m_{\rm DM}(=M_1), M_3, m_0, \lambda\}$ in the scotogenic model for $\Omega_{\rm DM}h^2 = 0.12 \pm 0.0009$ and $6 \times 10^{-14} \le {\rm BR}(\mu \rightarrow e \gamma) \le 4.2\times 10^{-13}$ in the case of $\beta > \beta_{\rm c}$ (PBH dominated era).}
\label{fig:PBHdom_beta_param}
\end{center}
\end{figure}

\begin{figure}[t]
\begin{center}
\includegraphics[scale=0.8]{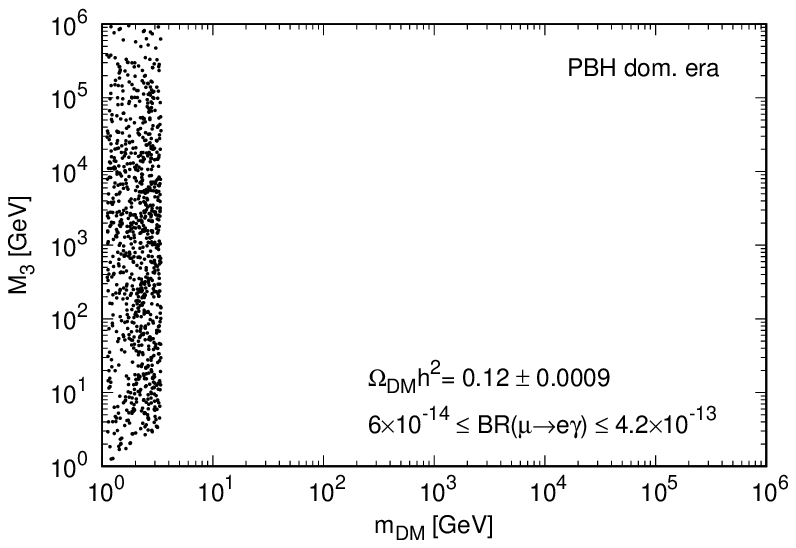}
\includegraphics[scale=0.8]{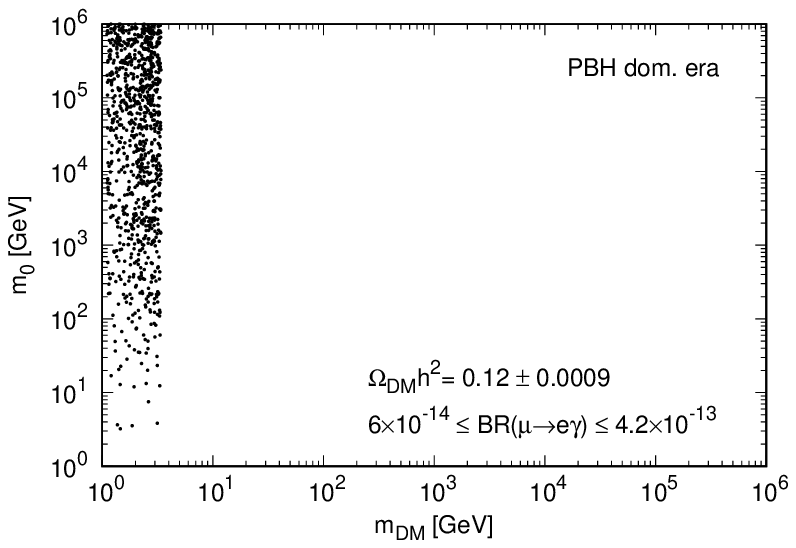}\\
\includegraphics[scale=0.8]{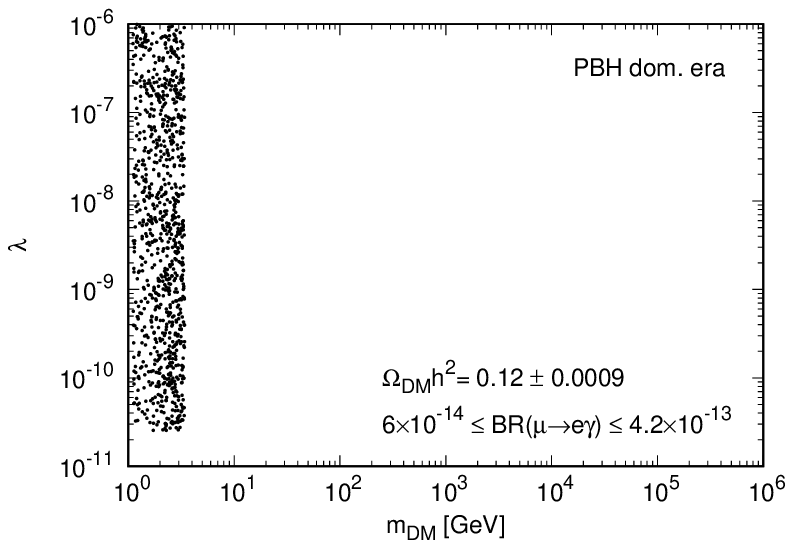}
\includegraphics[scale=0.8]{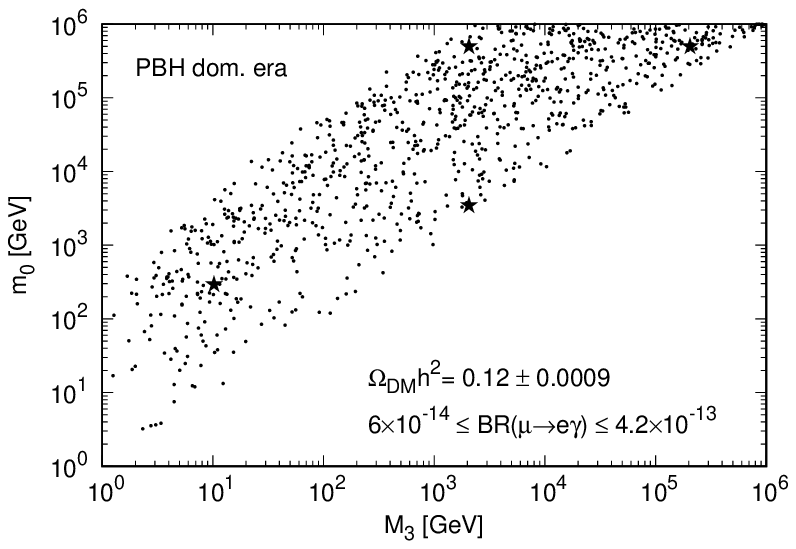}\\
\includegraphics[scale=0.8]{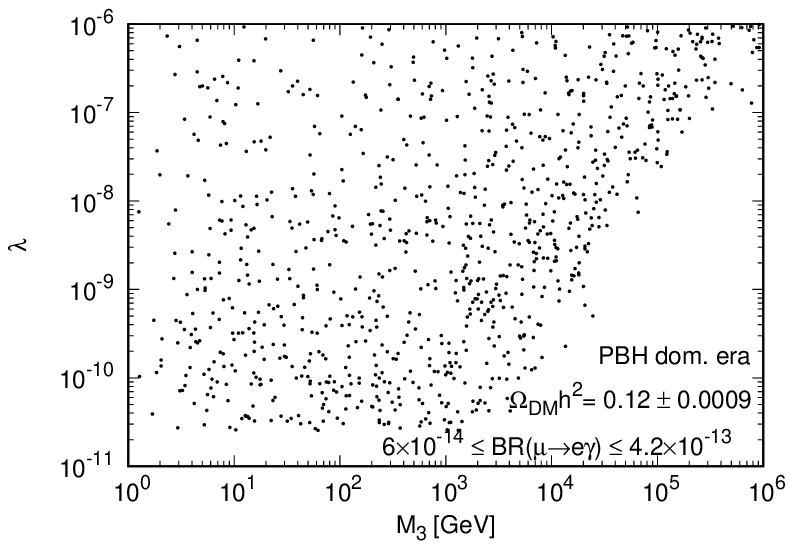}
\includegraphics[scale=0.8]{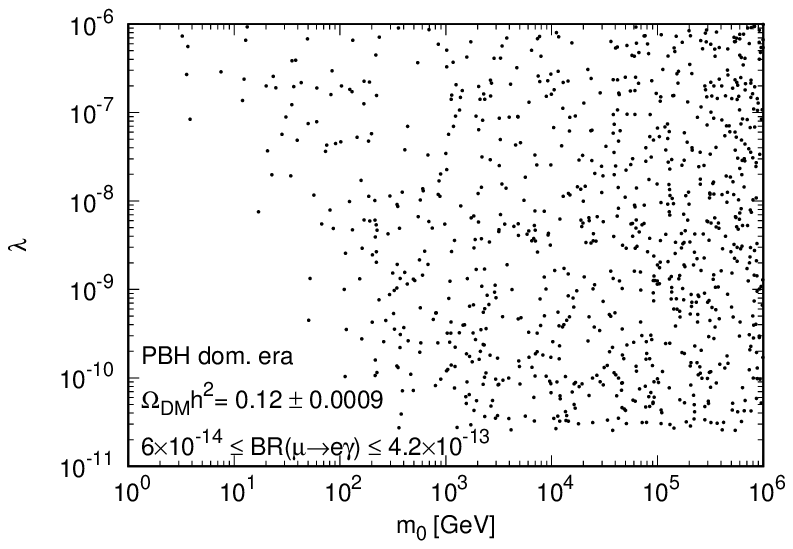}
\caption{Correlations between the allowed regions of 4 free parameters $\{m_{\rm DM}(=M_1), M_3, m_0, \lambda\}$ in the scotogenic model for $\Omega_{\rm DM}h^2 = 0.12 \pm 0.0009$ and $6 \times 10^{-14} \le {\rm BR}(\mu \rightarrow e \gamma) \le 4.2\times 10^{-13}$ in the case of $\beta > \beta_{\rm c}$ (PBH dominated era). The mark $\star$ denotes a benchmark case.}
\label{fig:PBHdom_socoto_params}
\end{center}
\end{figure}

\begin{figure}[t]
\begin{center}
\includegraphics[scale=0.8]{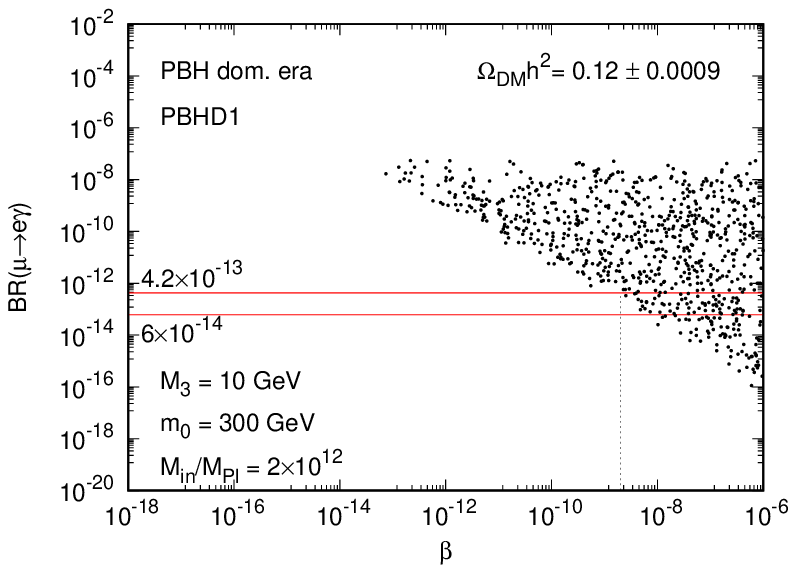}
\includegraphics[scale=0.8]{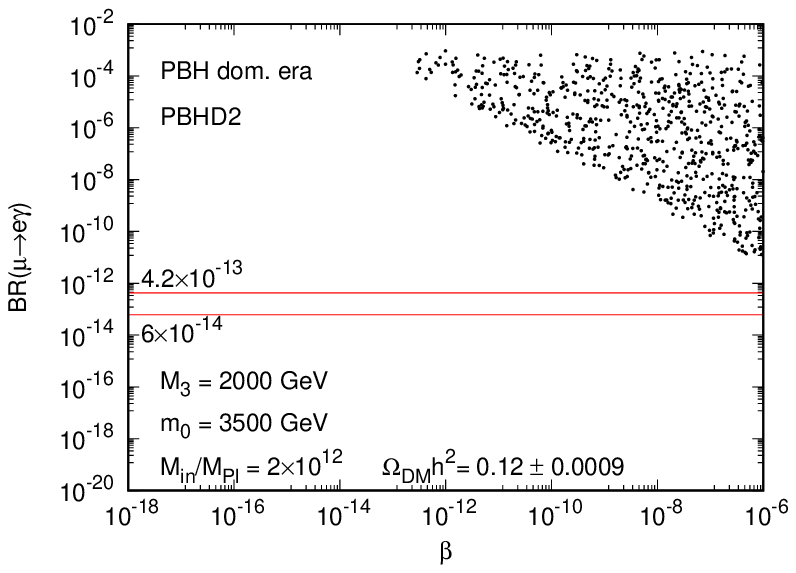}\\
\includegraphics[scale=0.8]{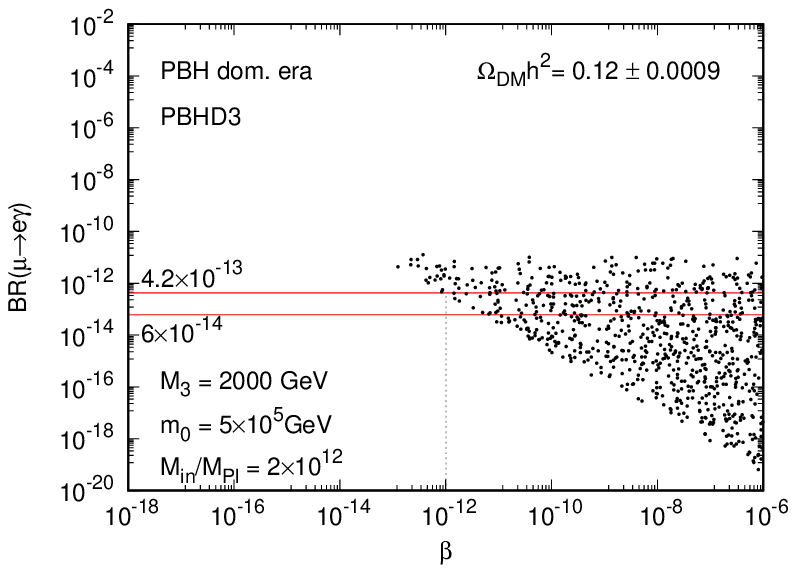}
\includegraphics[scale=0.8]{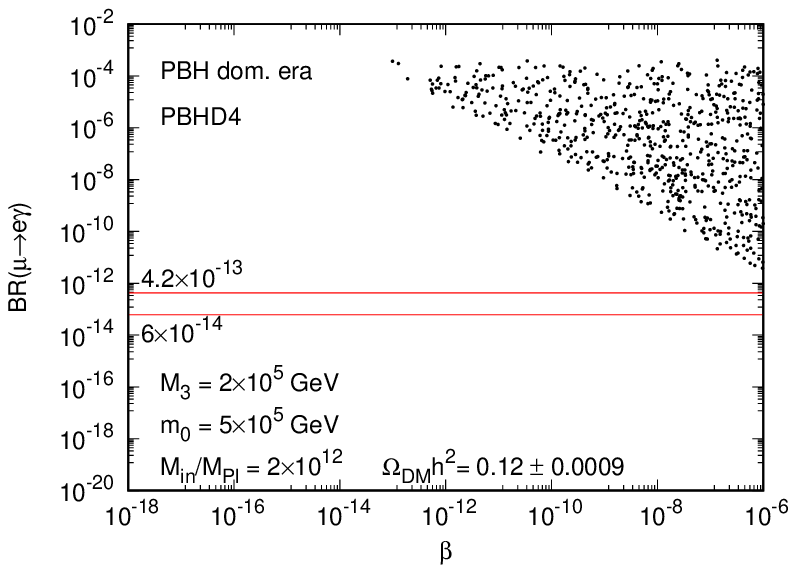}\\
\includegraphics[scale=0.8]{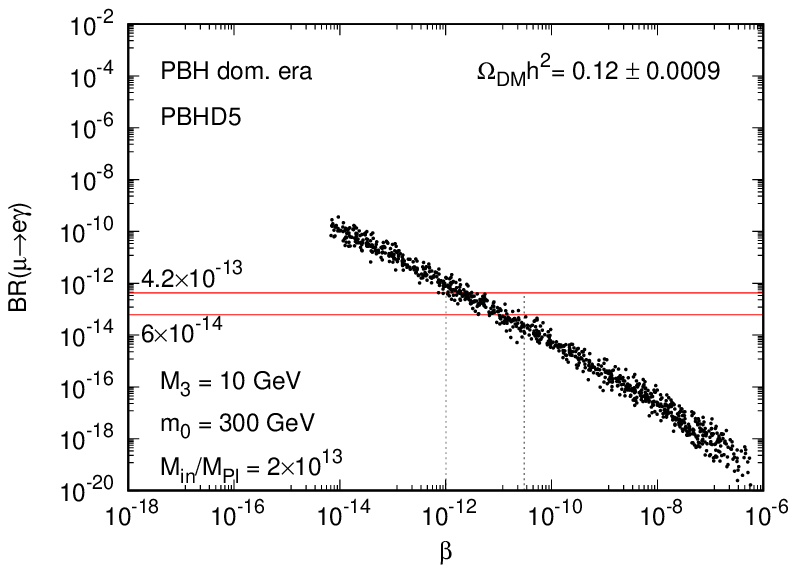}
\includegraphics[scale=0.8]{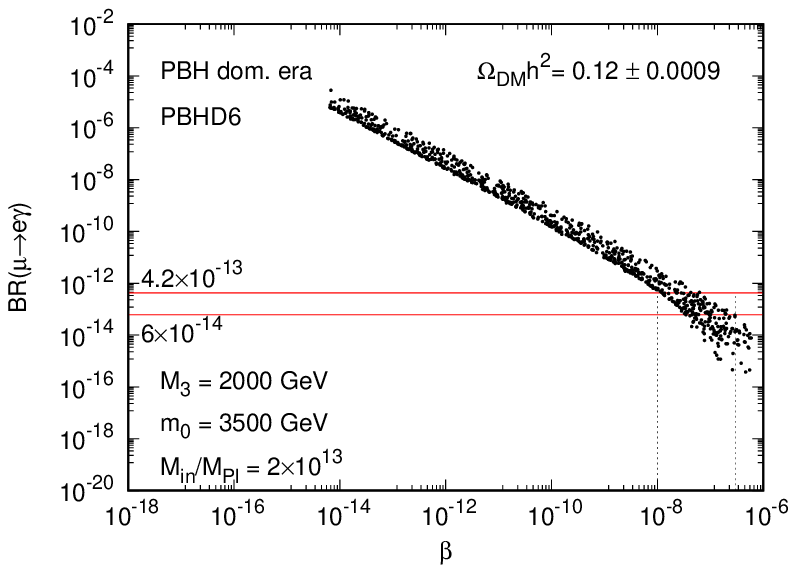}\\
\includegraphics[scale=0.8]{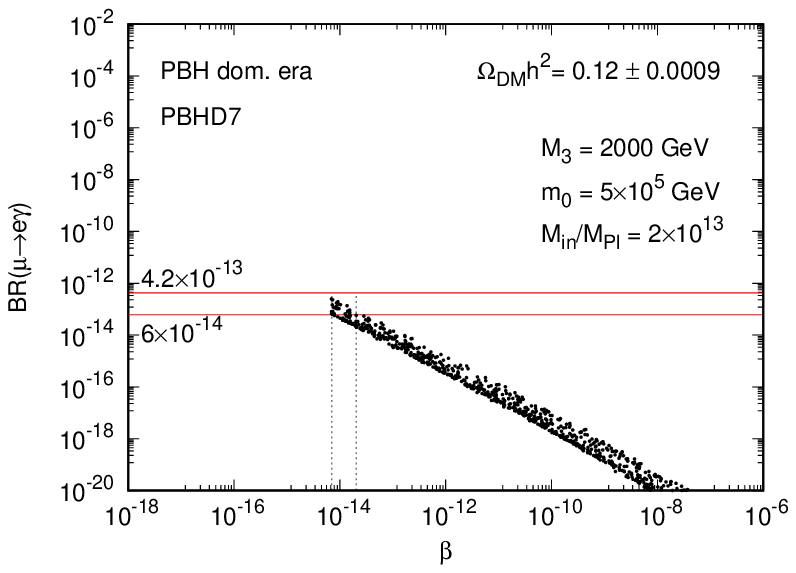}
\includegraphics[scale=0.8]{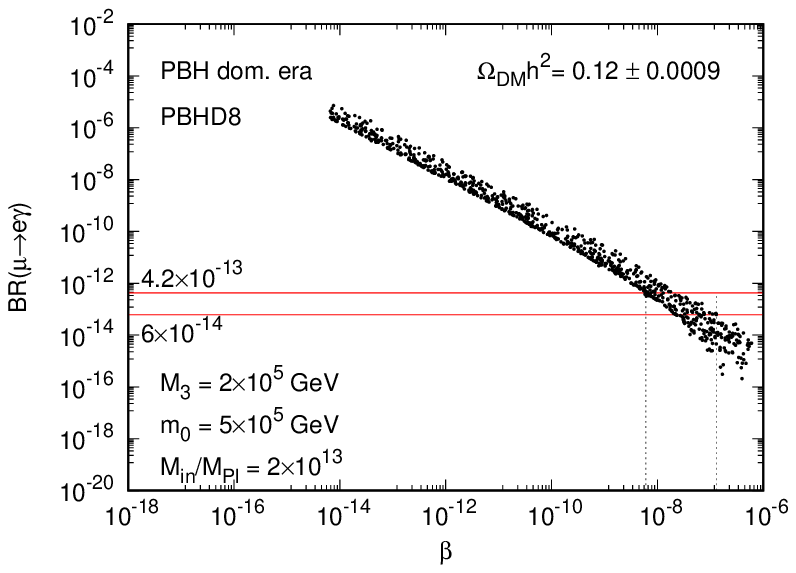}
\caption{Same of Fig. \ref{fig:PBHdom_Br_beta} but for the benchmark cases.}
\label{fig:PBHdom_Br_beta_BP}
\end{center}
\end{figure}

\begin{figure}[t]
\begin{center}
\includegraphics{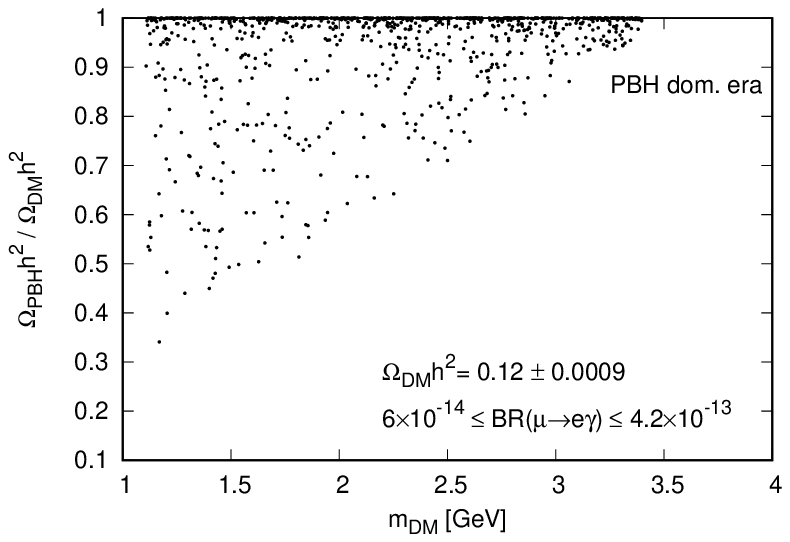}
\caption{Portion of the dark matter particles come from PBHs for $\Omega_{\rm DM}h^2 = 0.12 \pm 0.0009$ and $6 \times 10^{-14} \le {\rm BR}(\mu \rightarrow e \gamma) \le 4.2\times 10^{-13}$ in the case of $\beta > \beta_{\rm c}$ (PBH dominated era).}
\label{fig:PBHdom_OmegaPBHh2_OmegaDMh2}
\end{center}
\end{figure}

If the condition $\beta  > \beta_{\rm c}$ is satisfied, PBH evaporation occurs after the early equality time (PBH dominated era). In the PBH dominated era, the entropy production via the evaporation of PBHs leads to a dilution of  the freeze-out-origin scotogenic dark matter \cite{Fujita2014PRD,Hamdan2018MPLA,Masina2020EPJP,Baldes2020JCAP,KolbTurner1991}. The final relic abundance of scotogenic dark matter is obtained as
\begin{eqnarray}
\Omega_{\rm DM} h^2 = \alpha^{-1} \Omega_{\rm FO}h^2 + \Omega_{\rm PBH}h^2,
\end{eqnarray}
where $\alpha$ denotes the entropy boost factor. The factor $\alpha$ is the ratio of the entropy prior $S_{\rm before}$ and after  $S_{\rm after}$ the PBH evaporation, $\alpha (s_{\rm before} a_{\rm before} ^3) = s_{\rm after} a_{\rm after} ^3$, and is given by
\begin{eqnarray}
\alpha = \frac{S_{\rm after}}{S_{\rm before}} =\frac{Y_{\rm in}}{Y_{\rm evap}}, 
\label{Eq:alpha}
\end{eqnarray}
where
\begin{eqnarray}
Y_{\rm in} &=& \frac{n_{\rm BH}(t_{\rm in})}{s(t_{\rm in})} = \beta \frac{\rho_{\rm rad}(t_{\rm in})}{M_{\rm in}s(t_{\rm in})}, \nonumber \\
Y_{\rm evap} &=& \frac{n_{\rm BH}(t_{\rm evap})}{s(t_{\rm evap})} = \frac{\rho_{\rm rad}(t_{\rm evap})}{M_{\rm in}s(t_{\rm evap})}, 
\label{Eq:Yin_Yevap}
\end{eqnarray}
with 
\begin{eqnarray}
\rho_{\rm rad}(T) &=& \frac{\pi^2}{30} g_*(T)T^4, \nonumber \\
s(T) &=& \frac{2\pi^2}{45} g_{*s}(T)T^3.
\label{Eq:rho_rad_s}
\end{eqnarray}
where $g_{\ast s}$ is the relativistic effective degrees of freedom for entropy density. By combining Eqs. (\ref{Eq:alpha}), (\ref{Eq:Yin_Yevap}) and (\ref{Eq:rho_rad_s}), we obtain 
\begin{eqnarray}
\alpha = \beta \frac{g_\ast(t_{\rm in})}{g_\ast(t_{\rm evap})}\frac{g_{\ast s}(t_{\rm evap})}{g_{\ast s}(t_{\rm in})}\frac{T_{\rm in}}{T_{\rm evap}}.
\end{eqnarray}
According to the relation $g_\ast(T) \simeq g_{\ast s}(T)$ for high temperature, the entropy boost factor becomes the ratio of the initial density of PBH $\beta$ and critical density $\beta_{\rm c}$:
\begin{eqnarray}
\alpha = \frac{\beta}{\beta_{\rm c}}.
\end{eqnarray}

From the relation of $\Omega_{\rm DM} h^2 = \alpha^{-1} \Omega_{\rm FO}h^2 + \Omega_{\rm PBH}h^2$, at least, the relic abundance of dark matter via PBH evaporation, $\Omega_{\rm PBH}h^2$ should be less than the observed relic abundance $\Omega_{\rm DM} h^2 = 0.12\pm 0.0009$. 

First, we estimate the allowed region of the dark matter mass $m_{\rm DM}$ without relic abundance via freeze-out mechanism. Figure \ref{fig:PBHdom_omegaPBH_mDM} shows the dependence of relic abundance of PBH-origin scotogenic dark matter $\Omega_{\rm PBH}h^2$ on dark matter mass $m_{\rm DM}$ in the case of $\beta > \beta_{\rm c}$. The horizontal line shows the observed relic abundance of dark matter. We require $M_{\rm in}/M_{\rm Pl} \gtrsim 1\times 10^{10}$ for $m_{\rm DM} \simeq 1$ GeV. The comment for this requirement will be addressed later. With this requirement, the following regions of scotogenic dark matter mass 
\begin{eqnarray}
1.1 {\rm GeV} \le m_{\rm DM} \le 3.4 {\rm GeV},
\end{eqnarray}
and initial PBH mass
\begin{eqnarray}
2 \times 10^{12}  \lesssim M_{\rm in}/M_{\rm Pl} \lesssim 2 \times 10^{13},
\end{eqnarray}
are relevant for $\beta > \beta_{\rm c}$.

For $m_{\rm DM} = 1.1  - 3.4 $ GeV, we have
\begin{eqnarray}
\frac{M_{\rm in}}{M_{\rm Pl}} \gtrsim 
\begin{cases}
1.88 \times 10^{12} & (m_{\rm DM} = 1.1 {\rm GeV})\\ 
8.85 \times 10^{11} & (m_{\rm DM} = 3.4 {\rm GeV})
\end{cases}.
\end{eqnarray}
The Hawking temperature at the PBH formation time to be
\begin{eqnarray}
T_{\rm BH}^{\rm in}= 
\begin{cases}
2.6 \times 10^{5} {\rm GeV} & (\frac{M_{\rm in}}{M_{\rm Pl}} =1.88\times 10^{12}) \\ 
5.5 \times 10^{5} {\rm GeV} & (\frac{M_{\rm in}}{M_{\rm Pl}} = 8.85 \times 10^{11}) 
\end{cases}.
\end{eqnarray}
Thus, the relation $T_{\rm BH}^{\rm in} > m_{\rm DM}$ is satisfied in our setup. The relic abundance of PBH-origin scotogenic dark matter is obtained as
\begin{eqnarray}
\Omega_{\rm PBH}  h^2 &\simeq&  1.09 \times 10^7  \left(\frac{g_*(T_{\rm BH})}{106.75} \right)^{1/4}   \frac{3}{4} \frac{g_{\rm DM}  }{g_*(T_{\rm BH})}  \left( \frac{m_{\rm DM}}{{\rm GeV}} \right) \left( \frac{M_{\rm Pl}}{M_{\rm in}} \right)^{1/2},
\label{Eq:OmegaPBHh2_MD_T>m} 
\end{eqnarray}
for $T_{\rm BH}^{\rm in} > m_{\rm DM}$ \cite{Fujita2014PRD,Hamdan2018MPLA,Masina2020EPJP,Baldes2020JCAP}.

Now, we include the relic abundance of the dark matter via freeze-out mechanism to our analysis. Figure \ref{fig:PBHdom_Br_beta} shows the correlations between the branching ratio ${\rm BR} (\mu \rightarrow e \gamma)$ and initial density of PBHs $\beta$ for $\Omega_{\rm DM} h^2 = 0.12 \pm 0.0009$ and $1.1 {\rm GeV} \le m_{\rm DM} \le 3.4 {\rm GeV}$ in the case of $\beta > \beta_{\rm c}$. The upper horizontal line shows the current observed upper limit of ${\rm BR}(\mu\rightarrow e\gamma)$. The lower horizontal line shows the expected sensitivity of the future MEG II experiment. From Fig. \ref{fig:PBHdom_Br_beta}, the initial density of PBHs should be 
\begin{eqnarray}
\beta \gtrsim 8 \times 10^{-15}, 
\end{eqnarray}
for $\Omega_{\rm DM} h^2 = 0.12 \pm 0.0009$ in the case of  $\beta > \beta_{\rm c}$ with scotogenic dark matter.

Figure \ref{fig:PBHdom_beta_param} shows the correlations between the initial density of PBHs $\beta$ and 4 free parameters $\{m_{\rm DM}(=M_1), M_3, m_0, \lambda\}$ in the scotogenic model for  $\Omega_{\rm DM}h^2 = 0.12 \pm 0.0009$ and $6 \times 10^{-14} \le {\rm BR}(\mu\rightarrow e\gamma) \le 4.2\times 10^{-13}$ in the case of $\beta > \beta_{\rm c}$. From Fig. \ref{fig:PBHdom_beta_param}, if $\mu \rightarrow e \gamma$ process is observed in the MEG II experiment, the allowed regions of the 4 free parameters to be:
\begin{eqnarray}
1.1  \lesssim &m_{\rm DM} \ [{\rm GeV}]&  \lesssim 3.4, \nonumber \\
1.3  \lesssim &M_3 \ [{\rm GeV}]& \lesssim 1 \times 10^{6}, \nonumber \\
3.6 \lesssim &m_0 \ [{\rm GeV}]&  \lesssim  1 \times 10^{6}, \nonumber \\
2.5 \times 10^{-11} \lesssim &\lambda&  \lesssim 1 \times 10^{-6}, 
\end{eqnarray}
in the case of $\beta > \beta_{\rm c}$. The allowed region of the DM mass is narrow for $\beta > \beta_{\rm c}$. In the case of $\beta > \beta_{\rm c}$ (PBH  dominated era), the relic abundance via PBH evaporation may be more dominant than the relic abundance via freeze-out mechanism in observed relic abundance\cite{Gondolo2020PRD}. In this case, the allowed region of dark matter mass should be narrow to satisfy the condition of $\Omega_{\rm PBH}h^2 \le \Omega_{\rm DM}h^2 = 0.12$ for $M_{\rm in}/M_{\rm Pl} \le 2\times 10^{10}$ as shown in Fig.\ref{fig:PBHdom_omegaPBH_mDM}.

Figure \ref{fig:PBHdom_socoto_params} shows the correlations between the allowed regions of 4 free parameters $\{m_{\rm DM}(=M_1), M_3, m_0, \lambda\}$ in the scotogenic model in the case of $\beta > \beta_{\rm c}$. The mark $\star$ denotes a benchmark case. We pick up the following 4 benchmark cases in the case or $\beta > \beta_{\rm c}$ (PBH dominated era (PBHD)):
\begin{description}
\item[PBHD1:] $\{M_3, m_0\} =\{10, 300\}$ GeV as a light masses set.
\item[PBHD2:] $\{M_3, m_0\} =\{2000, 3500\}$ GeV as a middle masses set. 
\item[PBHD3:] $\{M_3, m_0\} =\{2000, 5\times 10^{5}\}$ GeV as an other middle masses set.
\item[PBHD4:] $\{M_3, m_0\} =\{2\times 10^{5}, 5\times 10^{5}\}$ GeV as a heavy masses set.
\end{description}
for $M_{\rm in}/M_{\rm Pl} = 2 \times 10^{12}$. In addition, we consider more 4 benchmark cases, {\bf PBHD5, PBHD6, PBHD7, PBHD8}, with same $\{M_3, m_0\}$ sets in PBHD1, PBHD2, PBHD3, PBHD4, respectively, for $M_{\rm in}/M_{\rm Pl} = 2 \times 10^{13}$. Since the allowed region of the dark matter mass is narrow, we vary the dark matter mass as $m_{\rm DM} =1.1 - 3.4$ GeV in these 8 benchmark cases. In addition, to avoid the large number of benchmark case, we have distinguished the benchmark cases by 3 parameter $\{M_3, m_0\}$ and 
$M_{\rm in}/M_{\rm Pl}$ without $\lambda$.

Figure \ref{fig:PBHdom_Br_beta_BP} shows the same of Fig. \ref{fig:PBHdom_Br_beta} but for the benchmark cases. We observe that if the lepton flavor violating $\mu \rightarrow e \gamma$ processes is observed in the MEG II experiment, $4.2 \times 10^{-13} \lesssim {\rm BR}(\mu\rightarrow e\gamma) \lesssim 6 \times 10^{-14}$, the initial density of PBH should be constrained for each benchmark case as follows:
\begin{description}
\item[PBHD1:] $2 \times 10^{-9} \lesssim \beta$.
\item[PBHD2:] No constraint.
\item[PBHD3:] $1 \times 10^{-12} \lesssim \beta$.
\item[PBHD4:] No constraint.
\item[PBHD5:] $1 \times 10^{-12} \lesssim \beta \lesssim 3 \times 10^{-11}$.
\item[PBHD6:] $1 \times 10^{-8} \lesssim \beta \lesssim 3 \times 10^{-7}$.
\item[PBHD7:] $7 \times 10^{-15} \lesssim \beta \lesssim 2 \times 10^{-14}$.
\item[PBHD8:] $6 \times 10^{-9} \lesssim \beta \lesssim 1 \times 10^{-7}$.
\end{description}

Now, we would like to comment about our requirement of $M_{\rm in}/M_{\rm Pl} \gtrsim 1 \times 10^{10}$ for $m_{\rm DM} \simeq 1$ GeV. Baldes et al. show that if the all relic abundance comes from PBH evaporation, $\Omega_{\rm PBH}h^2 = \Omega_{\rm DM}h^2$, for $m_{\rm DM} \simeq 1$ GeV, PBHs with mass $M_{\rm in}/M_{\rm Pl} \lesssim 1 \times 10^{10}$ are not allowed by the conservative Lyman-$\alpha$ bound for warm dark matter mass $m_{\rm WDM} > 3$ keV \cite{Baldes2020JCAP}. Figure \ref{fig:PBHdom_OmegaPBHh2_OmegaDMh2} shows the portion of the dark matter particles come from PBHs for $\Omega_{\rm DM}h^2 = 0.12 \pm 0.0009$ and $6 \times 10^{-14} \le {\rm BR}(\mu \rightarrow e \gamma) \le 4.2\times 10^{-13}$ in the case of $\beta > \beta_{\rm c}$. From Fig. \ref{fig:PBHdom_OmegaPBHh2_OmegaDMh2}, we observe the almost all relic abundance may be caused via PBH evaporation in some specific parameter sets in the scoogenic model. In this case, the PBHs with mass $M_{\rm in}/M_{\rm Pl} \lesssim 1 \times 10^{10}$ for $m_{\rm DM} \simeq 1$ GeV are not allowed by the warm dark matter constraints. Thus, we have conservatively required the condition $M_{\rm in}/M_{\rm Pl} \gtrsim 1 \times 10^{10}$ for $m_{\rm DM} \simeq 1$ GeV in our analysis. 

Finally, we would like to address the effect of having an extended dark sector may have on the contribution to the final relic abundance of the dark matter. Not only the standard model particles and the dark matter particle but also the heavier Majorana particles, $N_2, N_3$ and the new scalar $\eta$ will be created via PBH evaporation. In our setup, the Hawking temperature of the PBHs is enough to be able to produce such particles.  These particles will  decay into dark matter $N_1$ via $N_3 \rightarrow \ell^\pm \ell^\mp N_{1,2}$, $N_2 \rightarrow \ell^\pm \ell^\mp N_1$ and $\eta \rightarrow \ell N_{1,2,3}$ \cite{Ma2006PRD}, and increase the contribution on the relic density of the dark matter. This remarkable effect of the decaying PBH-origin heavy particles to the relic abundance is studied in general scheme with simple and predictive particle model by Cheek et al. very recently \cite{Cheek2021arXiv2}. If we include this effect, the results in this paper may be modified. 

For example, $N_2$ decays into dark matter $N_1$ and a pair of leptons with the decay rate \cite{Molinaro2014JCAP,Baumholzer2020JHEP} 
\begin{eqnarray}
\Gamma (N_2 \rightarrow \ell_\alpha \ell_\beta N_1) &=& \frac{m_{N_2}^5}{6144 \pi^3 M^4}  \left( |Y_{1\beta}|^2 |Y_{2\alpha}|^2  + |Y_{1\alpha}|^2  |Y_{2\beta}|^2 \right)
\end{eqnarray}
where $M$ denotes the mass of the scalar particle that is exchanged in the process and $\alpha$ and $\beta$ denote the flavor of final state leptons. This decay gives a contribution to the dark matter relic abundance of the form $\Omega_{N_2\rightarrow N_1}h^2 = m_{N_1}/m_{N_2} \Omega_{N_2}h^2$ where $ \Omega_{N_2}h^2$ may be interpreted as the amount of $N_2$ via PBH evaporations \cite{Baumholzer2020JHEP}. This contribution can be constrained from the effective number of neutrinos in the early Universe and the condition
\begin{eqnarray}
\frac{\Omega_{N_2\rightarrow N_1}h^2}{\Omega h^2} \lesssim 0.2 \%,
\end{eqnarray}
should be satisfied for consistency with cosmological observations with typical masses of the particle in the scotogenic model \cite{Baumholzer2020JHEP}. From this knowledge, we may expect that the effect of the decaying PBH-origin heavy particles to the dark matter final abundance does not disturb drastically the conclusion of this paper; however, including this effect for more precise analysis may give us the valuable results as shown by Cheek et al. \cite{Cheek2021arXiv2}. In this paper, we would like to report our results as quasi-precise analysis and to allow us to ignore this effect. We intend to study with this important effect in our future study.

\section{Summary\label{sec:summary}}
In this paper, we have shown the correlations between the initial density of PBHs and the branching ratio of $\mu \rightarrow e \gamma$ with scotogenic dark matter. Since the scotogenic model can account for dark matter candidates and predict the lepton flavor violating processes simultaneously, and the scotogenic dark matter can be also produced by Hawking radiation of PBH, the initial density of the PBHs and the branching ratio of $\mu \rightarrow e \gamma$ is related via scotogenic dark matter. 

It turned out that if the lepton flavor violating $\mu \rightarrow e \gamma$ process is observed in the MEG II experiment, the initial density of primordial black holes (PBHs) can be constrained with the scotogenic dark matter. As a benchmark case, if the PBH evaporation occurs in the radiation dominated era, the initial density may be $\beta \lesssim 3 \times 10^{-16}$ for $\mathcal{O}$ (TeV) scale dark sector in the scotogenic model. As an other benchmark case, if PBHs evaporate in the PBH dominated era, the initial density may be $1 \times 10^{-8} \lesssim \beta \lesssim 3 \times 10^{-7}$ for $\mathcal{O}$ (GeV) scale dark matter with other $\mathcal{O}$ (TeV) scale particles in the scotogenic model.

Since the physics run in the MEG II experiment is about to be started, the predictions in this study may be tested in less than 5 years.




%

\vspace{0.2cm}
\noindent




\end{document}